\documentclass[12pt]{article}
\usepackage{amsmath,amssymb,theorem,cite,epsfig,url,psfrag,eepic,mathtools}
\advance \topmargin by -\headheight
\advance \topmargin by -\headsep     
\evensidemargin \oddsidemargin
\def\Maketitle{{\def\newpage{}\maketitle}}
\makeatletter
\def\Appendix{\appendix
  \def\@seccntformat##1{Appendix~\csname the##1\endcsname.~~}}
\makeatother
\makeatletter
\@addtoreset{equation}{section}

\makeatother
\theorembodyfont{\sffamily}
\newtheorem{theorem}{Theorem}[section]
\newtheorem{lemma}[theorem]{Lemma}
\newtheorem{Proposition}[theorem]{Proposition}
\newtheorem{corollary}[theorem]{Corollary}

\newtheorem{identity}[theorem]{Identity}

\def\jac{\mathop{\bf J}\nolimits}
\def\Xint#1{\mathchoice
{\XXint\displaystyle\textstyle{#1}}%
{\XXint\textstyle\scriptstyle{#1}}%
{\XXint\scriptstyle\scriptscriptstyle{#1}}%
{\XXint\scriptscriptstyle\scriptscriptstyle{#1}}%
\!\int}
\def\XXint#1#2#3{{\setbox0=\hbox{$#1{#2#3}{\int}$}
\vcenter{\hbox{$#2#3$}}\kern-.5\wd0}}

\def\dashint{\Xint-}

\def\ñth{{\textrm{\,ñth}}}
\begin{document}
\title{\textbf{On combinatorial expansion of the conformal blocks\\  arising from AGT conjecture}\vspace*{1cm}}
\author{V.~A.~Alba$^{1,2,3,4}$, V.~A.~Fateev$^{1,5}$, A.~V.~Litvinov$^{1}$ and G.~M.~Tarnopolskiy$^{1,3}$\vspace*{10pt}\\[\medskipamount]
$^1$~\parbox[t]{0.85\textwidth}{\normalsize\it\raggedright
Landau Institute for Theoretical Physics,
142432 Chernogolovka, Russia}\vspace*{5pt}\\[\medskipamount]
$^2$~\parbox[t]{0.85\textwidth}{\normalsize\it\raggedright
Bogolyubov Institute for Theoretical Physics NASU, 03680 Kyiv, Ukraine}\vspace*{5pt}\\[\medskipamount]
\hspace*{4pt}$^3$~\parbox[t]{0.85\textwidth}{\normalsize\it\raggedright
Department of General and Applied Physics, Moscow Institute of Physics and Technology, 
141700 Dolgoprudny, Russia}
\vspace*{5pt}\\[\medskipamount]
$^4$~\parbox[t]{0.85\textwidth}{\normalsize\it\raggedright
Institute for Theoretical and Experimental Physics, 117218 Moscow, Russia}\vspace*{5pt}\\
$^5$~\parbox[t]{0.85\textwidth}{\normalsize\it\raggedright
Laboratoire de Physique Th\'eorique et Astroparticules, UMR5207 CNRS-UM2, Universit\'e
Montpellier~II, 34095 Montpellier, France}}
\date{}
\rightline{\texttt{\today}}
\Maketitle
\begin{abstract}\vspace*{10pt}
In their recent paper \cite{Alday:2009aq} Alday, Gaiotto and Tachikawa proposed a relation between $\mathcal{N}=2$ four-dimensional supersymmetric gauge theories and two-dimensional conformal field theories. As part of their conjecture they gave an explicit combinatorial formula for the expansion of the conformal blocks inspired from the exact form of instanton part of the Nekrasov partition function. In this paper we study the origin of such an expansion from a CFT point of view. We consider the algebra $\mathcal{A}=\text{\sf Vir}\otimes\mathcal{H}$ which is  the tensor product of mutually commuting  Virasoro and Heisenberg algebras and discover the special orthogonal basis of states in the highest weight representations of $\mathcal{A}$. The matrix elements of primary fields in this basis have a very simple factorized form and coincide with the function called $Z_{\text{\sf{bif}}}$ appearing in the instanton counting literature. Having such a simple basis, the problem of computation of the conformal blocks simplifies drastically and can be shown to lead to the expansion proposed in \cite{Alday:2009aq}. We found that this basis diagonalizes an infinite system of commuting Integrals of Motion related to Benjamin-Ono integrable hierarchy.
\end{abstract}
\tableofcontents
\section{Introduction}
The bootstrap approach to two-dimensional conformal field theory was suggested in the seminal paper \cite{Belavin:1984vu} by Belavin, Polyakov and Zamolodchikov. Their main idea was to use simultaneously the conformal symmetry of the theory and the hypothesis about the operator algebra of local fields \cite{Polyakov:1974gs}. Namely, if one supposes the existence of the complete set of local fields  $\{\mathcal{O}_{k}(\xi)\}$ in the theory  then the completeness of this set is equivalent to the operator algebra (OPE)
\begin{equation}\label{OPE}
   \mathcal{O}_{i}(\xi)\mathcal{O}_{j}(0)=\sum_{k}C_{ij}^{k}(\xi)\mathcal{O}_{k}(0).
\end{equation}
The structure constants $C_{ij}^{k}(\xi)$ are some single-valued functions which are subject to the  infinite system of equations following from the condition of the associativity of the operator algebra \eqref{OPE}. In general, this system of equations is rather complicated to be solved exactly and the OPE would be of limited usefulness. However, in two-dimensional CFT one can proceed further since the conformal group is infinite-dimensional in this case and it implies the strong restriction on  the possible form of the structure constants $C_{ij}^{k}(\xi)$.
One can show that the complete set of fields $\{\mathcal{O}_{k}(\xi)\}$ decomposes in this case into direct sum of conformal families
\begin{equation}
    \{\mathcal{O}_{k}(\xi)\}=\sum_{n}\,[\Phi_{n}].
\end{equation}
The ancestor of each family $\Phi_{n}$ is called primary field. It transforms in the simple manner
\begin{equation}
    \Phi_{n}(z,\bar{z})\longrightarrow\left(\frac{dw}{dz}\right)^{\Delta_{n}}\left(\frac{d\bar{w}}{d\bar{z}}\right)^{\bar{\Delta}_{n}}
    \Phi_{n}(w,\bar{w}),
\end{equation}
under the conformal transformations
\begin{equation*}
     z\rightarrow w(z),\quad \bar{z}\rightarrow\bar{w}(\bar{z}).
\end{equation*}
The quantum numbers $\Delta_{n}$ and $\bar{\Delta}_{n}$ are called conformal dimensions. Other representatives of the conformal family $[\Phi_{n}]$ are usually referred as descendant fields. Their conformal dimensions constitute an infinite integer sequence 
\begin{equation*}
    \Delta_{n}^{(k)}=\Delta_{n}+k,\qquad 
    \bar{\Delta}_{n}^{(\bar{k})}=\bar{\Delta}_{n}+\bar{k},
\end{equation*}
and each conformal family corresponds to some highest weight representation of the conformal group. In two dimensions the conformal group is the  tensor product of holomorphic and antiholomorphic Virasoro algebras 
\begin{equation}\label{Virasoro}
   \begin{aligned}
   &[L_{n},L_{m}]=(n-m)L_{n+m}+\frac{c}{12}(n^{3}-n)\,\delta_{n+m,0},\\
   &[\bar{L}_{n},\bar{L}_{m}]=(n-m)\bar{L}_{n+m}+\frac{c}{12}(n^{3}-n)\,\delta_{n+m,0},
   \end{aligned}
\end{equation}
and hence the conformal family is a tensor product $[\Phi_{n}]=\pi_{n}\otimes\bar{\pi}_{n}$ of two Verma modules over the Virasoro algebra. The parameter $c$ in \eqref{Virasoro} is an important characteristic of CFT called central charge. Moreover, it can be shown that all the structure constants $C_{ij}^{k}(\xi)$ can be computed in terms of structure constants $\mathbb{C}_{ij}^{k}$ of primary fields \cite{Belavin:1984vu}.

This simple structure of the space of fields in two-dimensional CFT leads  to the introduction of the notion of the \emph{conformal blocks}. They represent holomorphic contributions to the  multi-point correlation function of primary fields 
\begin{equation}\label{n-point}
    \langle\Phi_{1}(z_{1},\bar{z}_{1})\dots\Phi_{n}(z_{n},\bar{z}_{n})\rangle
\end{equation}
picking given conformal families $\tilde{\Delta}_{j}$ in the intermediate channels. It is convenient to represent the $n-$point conformal block by the following picture \cite{Belavin:1984vu}\footnote{Throughout this paper we will consider only conformal blocks on a sphere i.e. on a surface of genus $0$.}
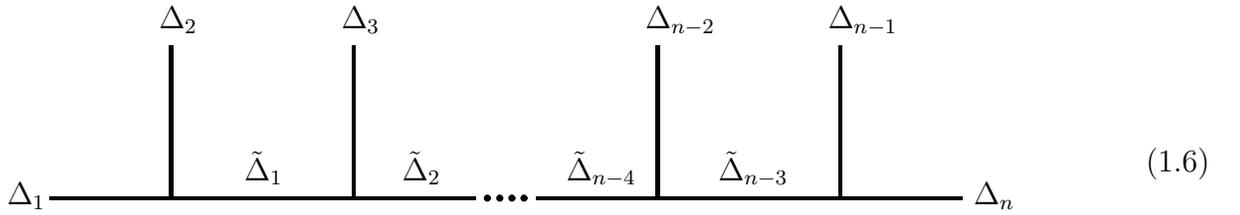
\begin{equation}\label{conformal-block}
    \begin{picture}(30,75)(160,10)
    \Thicklines
    \unitlength 2.3pt 
    \put(0,0){\line(1,0){70}}
    \put(72,0){\circle*{.7}}
    \put(74,0){\circle*{.7}}
    \put(76,0){\circle*{.7}}
    \put(78,0){\circle*{.7}}
    \put(80,0){\line(1,0){70}}
    \put(20,0){\line(0,1){25}}
    \put(50,0){\line(0,1){25}}
    \put(100,0){\line(0,1){25}}
    \put(130,0){\line(0,1){25}}
    \put(-7,-1){\mbox{$\Delta_{1}$}}
    \put(18,28){\mbox{$\Delta_{2}$}}
    \put(48,28){\mbox{$\Delta_{3}$}}
    \put(98,28){\mbox{$\Delta_{n-2}$}}
    \put(128,28){\mbox{$\Delta_{n-1}$}}
    \put(152,-1){\mbox{$\Delta_{n}$}}
    \put(32,3){\mbox{$\tilde{\Delta}_{1}$}}
    \put(58,3){\mbox{$\tilde{\Delta}_{2}$}}
    \put(85,3){\mbox{$\tilde{\Delta}_{n-4}$}}
    \put(110,3){\mbox{$\tilde{\Delta}_{n-3}$}}
    \end{picture}
    \vspace*{1cm}
\end{equation}
which encodes the way how the OPE in \eqref{n-point}  is performed. The $n-$point conformal block \eqref{conformal-block} is a function of holomorphic coordinates $z_{1},\dots,z_{n}$, of external dimensions $\Delta_{1},\dots,\Delta_{n}$, intermediate dimensions $\tilde{\Delta}_{1},\dots,\tilde{\Delta}_{n-3}$ and the central charge $c$. To be more specific it is convenient to use the projective invariance and fix $z_{1}=0$, $z_{n-1}=1$ and $z_{n}=\infty$. It is also convenient to choose
\begin{equation*}
    z_{i+1}=q_{i}q_{i+1}\dots q_{n-3}\quad\text{for}\quad1\leq i\leq n-3,
\end{equation*}
then the conformal block corresponding to the picture \eqref{conformal-block} is a power series expansion \cite{Belavin:1984vu}
\begin{equation}\label{conformal-block-explicit}
    \mathcal{F}(q|\Delta_{i},\tilde{\Delta}_{j},c)=
    1+\sum_{\vec{k}}q_{1}^{k_{1}}q_{2}^{k_{2}}\dots q_{n-3}^{k_{n-3}}\,\mathcal{F}_{\vec{k}}(\Delta_{i},\tilde{\Delta}_{j},c),
\end{equation}
where sum goes over all set of positive integers $\vec{k}=(k_{1},\dots,k_{n-3})$ and the  coefficients $\mathcal{F}_{\vec{k}}(\Delta_{i},\tilde{\Delta}_{j},c)$ are some rational functions of $\Delta_{i}$, $\tilde{\Delta}_{j}$ and the central charge $c$ which are completely determined (in principle) by the conformal symmetry \cite{Belavin:1984vu}. The idea is to ``cut'' all the intermediate necks of the conformal block \eqref{conformal-block} and to ``insert'' the complete set of states \cite{Moore:1988qv}.  Then the problem of computing the coefficients $\mathcal{F}_{\vec{k}}(\Delta_{i},\tilde{\Delta}_{j},c)$ in \eqref{conformal-block-explicit}  is equivalent to the problem of computing  the normalized matrix elements of arbitrary primary field  between CFT states  
\begin{equation}\label{Virasoro-matrix-element}
    \frac{\langle i|L_{k'_{1}}\dots L_{k'_{m}}\Phi_{k}(1)\,L_{-k_{n}}\dots L_{-k_{1}}|j\rangle}
    { \langle i|\Phi_{k}(1)|j\rangle}.
\end{equation}
For any two given sets of integers $(k_{1},\dots,k_{n})$ and $(k'_{1},\dots,k'_{m})$ it is a matter of algebraic manipulations to compute the matrix element \eqref{Virasoro-matrix-element} which is in fact some polynomial of conformal dimensions $\Delta_{i}$, $\Delta_{j}$ and $\Delta_{k}$. However, this computation becomes tedious for higher levels. It can be facilitated if one distinguishes between ``conformal fields'' and ``derivatives'' \cite{Yurov:1989yu}, but still it is rather difficult to find a closed form expression for all the matrix elements \eqref{Virasoro-matrix-element}. Explicit expressions  obtained by symbolic algebra program for lower levels show no signs of general pattern. So, it would be desirable to have more efficient algorithm for computation of the expansion \eqref{conformal-block-explicit}. One possibility known as the recursive procedure was suggested by Alyosha Zamolodchikov in \cite{Zamolodchikov:1985ie}. The method developed in \cite{Zamolodchikov:1985ie} was originally applied to four-point conformal block on a sphere and although the generalization of this method  for more general case seems to be possible there is no much available results in the existing literature\footnote{Recently, the recursive procedure of \cite{Zamolodchikov:1985ie} was generalized to  the case of one-point conformal block on a torus in \cite{Hadasz:2009db}.}. 

A renewed interest in conformal field theory appeared after the paper \cite{Alday:2009aq} of Alday, Gaiotto and Tachikawa where they proposed a relation between two-dimensional conformal field theories and $\mathcal{N}=2$ four-dimensional supersymmetric  gauge theories (it is usually referred as the AGT conjecture). In particular, they related  the $n-$point conformal block on a sphere \eqref{conformal-block-explicit} with the instanton part of the Nekrasov partition function \cite{Moore:1997dj,Nekrasov:2002qd,Nekrasov:2003rj} for the gauge theory with gauge group $U(2)_{1}\otimes\dots\otimes U(2)_{n-3}$ and with special matter content which is either in (anti-)fundamental representation of $U(2)_{1}$ or $U(2)_{n-3}$ or in bifundamental representation of $U(2)_{i}\otimes U(2)_{i+1}$ for $i=1,\dots,n-2$. Theories of such a type are usually called linear quiver gauge theories \cite{Douglas:1996sw,Gaiotto:2009gz,Gaiotto:2009we,Benini:2009gi}. In order to formulate the result of \cite{Alday:2009aq} we define the new function
\begin{equation}\label{conformal-block-refined}
   Z(q|\Delta_{i},\tilde{\Delta}_{j},c)\overset{\text{def}}{=}
    \prod_{k=1}^{n-3}\prod_{m=k}^{n-3}(1-q_{k}\dots q_{m})^{2\alpha_{k+1}(Q-\alpha_{m+2})}\,\,
    \mathcal{F}(q|\Delta_{i},\tilde{\Delta}_{j},c),
\end{equation}
where the parameters $\alpha_{k}$ and $Q$ were introduced to parametrize the external conformal dimensions $\Delta_{k}$ and the central charge $c$ as
\begin{equation}
     \Delta_{k}=\alpha_{k}(Q-\alpha_{k}),\quad c=1+6Q^{2},\qquad Q=b+\frac{1}{b}.
\end{equation}
It was conjectured in \cite{Alday:2009aq} that the function $Z(q|\Delta_{i},\tilde{\Delta}_{j},c)$ defined by \eqref{conformal-block-refined} possesses the nice expansion
\begin{equation}\label{conformal-block-refined-explicit}
    Z(q|\Delta_{i},\tilde{\Delta}_{j},c)=
    1+\sum_{\vec{k}}q_{1}^{k_{1}}q_{2}^{k_{2}}\dots q_{n-3}^{k_{n-3}}\,Z_{\vec{k}}(\Delta_{i},\tilde{\Delta}_{j},c),
\end{equation}
with the coefficients $Z_{\vec{k}}(\Delta_{i},\tilde{\Delta}_{j},c)$ having explicit combinatorial expression 
\begin{multline}\label{conformal-block-refined-explicit-coefficient}
   Z_{\vec{k}}(\Delta_{i},\tilde{\Delta}_{j},c)=
   \sum_{\vec{\lambda}_{1},\dots,\vec{\lambda}_{n-3}}
   Z_{\text{\sf{vec}}}(P_{1},\vec{\lambda}_{1})\dots Z_{\text{\sf{vec}}}(P_{n-3},\vec{\lambda}_{n-3})\times\\\times
   Z_{\text{\sf{bif}}}(\alpha_{2}|P,\varnothing;P_{1},\vec{\lambda}_{1})Z_{\text{\sf{bif}}}(\alpha_{3}|P_{1},\vec{\lambda}_{1};P_{2},\vec{\lambda}_{2})
   Z_{\text{\sf{bif}}}(\alpha_{4}|P_{2},\vec{\lambda}_{2};P_{3},\vec{\lambda}_{3})
   \times\dots\\\dots\times 
   Z_{\text{\sf{bif}}}(\alpha_{n-3}|P_{n-5},\vec{\lambda}_{n-5};P_{n-4},\vec{\lambda}_{n-4})
   Z_{\text{\sf{bif}}}(\alpha_{n-2}|P_{n-4},\vec{\lambda}_{n-4};P_{n-3},\vec{\lambda}_{n-3})
   Z_{\text{\sf{bif}}}(\alpha_{n-1}|P_{n-3},\vec{\lambda}_{n-3};\hat{P},\varnothing).
\end{multline}
The sum in \eqref{conformal-block-refined-explicit-coefficient} goes over the pairs $\vec{\lambda}=(\lambda_{1},\lambda_{2})$ of Young diagrams (bipartitions) such that $|\vec{\lambda}_{j}|=k_{j}$ where $|\vec{\lambda}_{j}|$ is the total number of boxes in the pair  $\vec{\lambda}_{j}$. The parameters $P$, $\hat{P}$ and $P_{j}$ in \eqref{conformal-block-refined-explicit-coefficient} are related with the external dimensions $\Delta_{1}$, $\Delta_{n}$ and  intermediate dimensions $\tilde{\Delta}_{j}$ by
\begin{equation*}
   \Delta_{1}=\frac{Q^{2}}{4}-P^{2},\quad
   \Delta_{n}=\frac{Q^{2}}{4}-\hat{P}^{2}\quad\text{and}\quad
   \tilde{\Delta}_{j}=\frac{Q^{2}}{4}-P_{j}^{2}.
\end{equation*}
The explicit form of the functions $Z_{\text{\sf{bif}}}$ and $Z_{\text{\sf{vec}}}$  were derived in \cite{Fucito:2004gi,Flume:2002az,Shadchin:2005cc}. The function $Z_{\text{\sf{bif}}}$ is given by
\begin{equation}\label{Zbif-def}
    Z_{\text{\sf{bif}}}(\alpha|P',\vec{\mu};P,\vec{\lambda})=\prod_{i,j=1}^{2}
    \prod_{s\in \lambda_{i}}\left(Q-E_{\lambda_{i},\mu_{j}}\bigl(P_{i}-P'_{j}\bigl|s\bigr)-\alpha\right)
    \prod_{t\in \mu_{j}}\left(E_{\mu_{j},\lambda_{i}}\bigl(P'_{j}-P_{i}\bigl|t\bigr)-\alpha\right),  
\end{equation}
where $\vec{P}=(P,-P)$, $\vec{P}'=(P',-P')$ and
\addtocounter{equation}{-1}
\begin{subequations}
\begin{equation}\label{E-def}
    E_{\lambda,\mu}\bigl(P\bigl|s\bigr)=P-b\,l_{\mu}(s)+b^{-1}(a_{\lambda}(s)+1).
\end{equation}
\end{subequations}
In \eqref{E-def} $a_{\lambda}(s)$ and $l_{\mu}(s)$ are correspondingly the arm length of the square $s$ in the partition $\lambda$ and the leg length of the square $s$ in the partition $\mu$. We choose the English convention to draw partitions $\lambda=(\lambda_{1}\geq\lambda_{2}\geq\dots)$. For example the partition $\lambda=(4,3,2,1,1)$ is drawn as follows
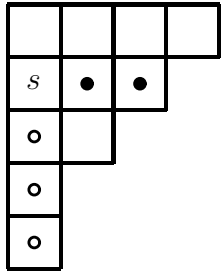
\begin{equation*}
  \begin{picture}(30,100)(20,20)
    \Thicklines
    \unitlength 2pt 
    \put(0,60){\line(1,0){40}}
    \put(0,50){\line(1,0){40}}
    \put(0,40){\line(1,0){30}}
    \put(0,30){\line(1,0){20}}
    \put(0,20){\line(1,0){10}}
    \put(0,10){\line(1,0){10}}
    \put(0,60){\line(0,-1){50}}
    \put(10,60){\line(0,-1){50}}
    \put(20,60){\line(0,-1){30}}
    \put(30,60){\line(0,-1){20}}
    \put(40,60){\line(0,-1){10}}
    \put(3.5,44){\mbox{$s$}}
    \put(15,45){\circle*{2}}
    \put(25,45){\circle*{2}}
    \put(5,35){\circle{2}}
    \put(5,25){\circle{2}}
    \put(5,15){\circle{2}}
    \end{picture}
\end{equation*} 
where the number of filled circles is equal to the arm length $a_{\lambda}(s)$ while the number of unfilled ones to the leg length $l_{\lambda}(s)$. We note that in \eqref{E-def} the arm and leg lengths are computed with respect to different partitions $\lambda$ and $\mu$ and the square $s$ always belongs to the partition $\lambda$ which is the first subscript of $E_{\lambda,\mu}\bigl(P\bigl|s\bigr)$. The function $Z_{\text{\sf{vec}}}$ is defined by
\begin{equation}\label{Zvec-def}
   Z_{\text{\sf{vec}}}(P,\vec{\lambda})=\frac{1}{Z_{\text{\sf{bif}}}(0|P,\vec{\lambda};P,\vec{\lambda})}.
\end{equation}
The AGT conjecture \cite{Alday:2009aq} attracts  an attention of both specialists in gauge theories and conformal field theory. In particular, we note that the combinatorial expansion \eqref{conformal-block-refined-explicit}-\eqref{conformal-block-refined-explicit-coefficient} was rather unexpected from CFT point of view. In a series of papers \cite{Mironov:2009qt,Marshakov:2009kj,Mironov:2009qn,Alba:2009ya,Fateev:2009aw,Hadasz:2010xp} this expansion was checked for some particular cases against CFT predictions. However,   general mathematically rigorous proof of \eqref{conformal-block-refined-explicit}-\eqref{conformal-block-refined-explicit-coefficient}  was lacking.

The factor in the r.h.s. in \eqref{conformal-block-refined} was called in \cite{Alday:2009aq}  the  ``$U(1)$ factor'' which presumably corresponds to the ``stripping off''  the $U(1)$ part from the Nekrasov partition function which is computed for $U(2)$ groups, rather than $SU(2)$. It looks natural to express the ``$U(1)$ factor'' in terms of correlation function of  chiral vertex operators of some free bosonic field.  We found that the introduction of the auxiliary bosonic field is not only the convenient way to represent the ``$U(1)$ factor'' but it plays a crucial r\^ole in  the whole construction. We consider  the algebra $\mathcal{A}=\text{\sf Vir}\otimes\mathcal{H}$ which is  the tensor product of Virasoro and Heisenberg algebras and construct the special orthogonal basis in the highest weight representations of this algebra. The matrix elements of primary fields between  any two states from this basis have particularly simple form which coincides with  $Z_{\text{\sf{bif}}}$ defined above. The norm of these states is equal to $1/Z_{\text{\sf{vec}}}$. It is clear that such a basis essentially leads to the expansion \eqref{conformal-block-refined-explicit}--\eqref{conformal-block-refined-explicit-coefficient}. Similar idea was proposed by Alday and Tachikawa in \cite{Alday:2010vg}. In this paper we prove the existence and uniqueness of such a basis and find underlaying quantum integrable system (the system of commuting Integrals of Motion diagonalized by this basis). 

The plan of the paper is the following. In section \ref{main-section} we formulate the Proposition \ref{MAIN-Proposition} which is the main result of our paper. Section \ref{Proof-section} is devoted to proof of Proposition  \ref{MAIN-Proposition}. In section \ref{Remarks} we make some concluding remarks. In appendix \ref{Selberg-proof} we compute Selberg integral with insertion of two Jack polynomials which is used in section \ref{Proof-section}. In appendix \ref{ME-reduction} we give a proof of the identities \eqref{Z-Z-relat} and \eqref{Z-Z-relat-2} used in section \ref{Proof-section}. In appendix \ref{IM} we discuss the system of quantum Integrals of Motion which appeared to be important in our problem. 
\section{Special basis of states in Vir$\otimes\mathcal{H}$}\label{main-section}
We consider the algebra $\mathcal{A}=\text{\sf Vir}\otimes\mathcal{H}$ which is  the tensor product of Virasoro and Heisenberg algebras with commutation relations
\begin{equation}
   \begin{aligned}
   &[L_{n},L_{m}]=(n-m)L_{n+m}+\frac{c}{12}(n^{3}-n)\,\delta_{n+m,0},\\
   &[a_{n},a_{m}]=\frac{n}{2}\,\delta_{n+m,0},\qquad [L_{n},a_{m}]=0.
   \end{aligned}
\end{equation} 
We will parametrize the central charge $c$ of Virasoro algebra in a Liouville manner as
\begin{equation}
   c=1+6Q^{2},\qquad\text{where}\quad Q=b+\frac{1}{b},
\end{equation}
and define the primary field $V_{\alpha}$ as
\begin{equation}\label{primary}
    V_{\alpha}\overset{\text{def}}{=}\mathcal{V}_{\alpha}\cdot V_{\alpha}^{\text{L}},
\end{equation}
where $V_{\alpha}^{\text{L}}$ is the primary field of Virasoro algebra with conformal dimension $\Delta(\alpha)=\alpha(Q-\alpha)$ and $\mathcal{V}_{\alpha}$ is a free exponential\footnote{We note that this ``strange'' form of the vertex operator \eqref{vertex} was suggested in somewhat different but related context by Carlsson and
Okounkov in \cite{Carlsson:2008fk}.}
\begin{equation}\label{vertex}
    \mathcal{V}_{\alpha}=e^{2(\alpha-Q)\varphi_{-}}e^{2\alpha\varphi_{+}},
\end{equation}
with $\varphi_{+}(z)=i\sum_{n>0}\frac{a_{n}}{n}z^{-n}$ and $\varphi_{-}(z)=i\sum_{n<0}\frac{a_{n}}{n}z^{-n}$. The commutation relations of the primary field $V_{\alpha}(z)$ with generators $L_{m}$ and $a_{n}$ can be summarized as
\begin{equation}\label{commutation-relations}
   \begin{aligned}
     &[L_{m},V_{\alpha}^{\text{L}}(z)]=\left(z^{m+1}\partial_{z}+(m+1)\Delta(\alpha)z^{m}\right)V_{\alpha}^{\text{L}}(z),\\
     &[a_{n},\mathcal{V}_{\alpha}(z)]=-i\alpha z^{n}\mathcal{V}_{\alpha}(z),&&\hspace*{-90pt}\text{for}\quad n<0,\\
     &[a_{n},\mathcal{V}_{\alpha}(z)]=i(Q-\alpha)z^{n}\mathcal{V}_{\alpha}(z),&&\hspace*{-90pt}\text{for}\quad n>0,\\
     &[L_{m},\mathcal{V}_{\alpha}(z)]=[a_{n},V_{\alpha}^{\text{L}}(z)]=0.
   \end{aligned}
\end{equation}
There is a natural basis in the space of states 
\begin{equation}\label{naive-basis}
    a_{-l_{m}}\dots a_{-l_{1}}L_{-k_{n}}\dots L_{-k_{1}}|P\rangle,\quad
    k_{1}\geq k_{2}\geq\dots\geq k_{n},\quad
    l_{1}\geq l_{2}\geq\dots\geq l_{m},
\end{equation}
where $P$ parametrizes the Virasoro conformal dimension as $\Delta(P)=\frac{Q^{2}}{4}-P^{2}$ and $|P\rangle$ is the vacuum state  which is defined by
\begin{equation*}
    L_{n}|P\rangle=a_{n}|P\rangle=0,\quad\text{for}\quad n>0,\qquad
    L_{0}|P\rangle=\Delta(P)|P\rangle,\qquad \langle P|P\rangle=1.
\end{equation*}
The matrix elements\footnote{Below we will hide the dependence on the insertion point when possible. So, in any expression below $V_{\alpha}$ means $V_{\alpha}(1)$.}
\begin{equation}\label{naive-matrix-element}
    \frac{\langle P'|L_{k'_{1}}\dots L_{k'_{n'}}a_{l'_{1}}\dots a_{l'_{m'}}\,
    V_{\alpha}(1)\,
    a_{-l_{m}}\dots a_{-l_{1}}L_{-k_{n}}\dots L_{-k_{1}}|P\rangle}
    {\langle P'|V_{\alpha}(1)|P\rangle},
\end{equation}
which are some polynomials in $\alpha$, $P$ and $P'$  can be computed using commutation rules \eqref{commutation-relations} and the explicit form of the coordinate dependence of the matrix element\footnote{The matrix element \eqref{naive-matrix-element} is recovered in the limit $z\rightarrow1$.}
\begin{equation}\label{form-of-3point-function}
   \langle P'|V^{L}_{\alpha}(z)|P\rangle\sim z^{P^{2}-P'^{2}-\Delta(\alpha)}.
\end{equation}

We note that the states   \eqref{naive-basis} are the eigenstates of the operator $L_{0}+2\sum_{k>0}a_{-k}a_{k}$ with the eigenvalues given by
\begin{equation}
   \Delta^{(k+l)}(P)\overset{\text{def}}{=}\Delta(P)+k+l,\quad k=\sum_{i=1}^{n}k_{i},\quad l=\sum_{j=1}^{m}l_{j}.
\end{equation}
For general values of the momenta $P$ the module $|P\rangle$ is irreducible and the number of states with given value of $\Delta^{(N)}(P)$ is equal to the number of pairs of Young diagrams $\vec{\lambda}=(\lambda_{1},\lambda_{2})$ with $|\vec{\lambda}|=N$. So, it is natural to define
\begin{equation}\label{naive-basis-2}
      \hat{a}_{-\lambda_{1}}\hat{L}_{-\lambda_{2}}|P\rangle\overset{\text{def}}{=}
      a_{-l_{m}}\dots a_{-l_{1}}L_{-k_{n}}\dots L_{-k_{1}}|P\rangle,
\end{equation}
where $\lambda_{1}=(l_{1},\dots,l_{m})$, $\lambda_{2}=(k_{1},\dots,k_{n})$.
We find it convenient to use different basis $|P\rangle_{\vec{\lambda}}$ instead of using naive one \eqref{naive-basis-2}
\begin{equation}\label{true-basis}
   |P\rangle_{\vec{\lambda}}=\sum_{|\vec{\mu}|=|\vec{\lambda}|}C_{\vec{\lambda}}^{\mu_{1},\mu_{2}}(P)\,
    \hat{a}_{-\mu_{1}}\hat{L}_{-\mu_{2}}|P\rangle,
\end{equation}
where the sum goes over all pairs  $\vec{\mu}=(\mu_{1},\mu_{2})$  of Young diagrams such that $|\vec{\mu}|=|\vec{\lambda}|$ and $C_{\vec{\lambda}}^{\mu_{1},\mu_{2}}(P)$ are some unknown coefficients. The meaning of pairs $\vec{\lambda}$ in \eqref{true-basis} is different from one in  \eqref{naive-basis-2} and will become clear below. 
We stress that the conjugation in the algebra $\mathcal{A}$ is defined as
\begin{equation}\label{conjugation}
    \left(L_{-k_{n}}\dots L_{-k_{1}}\right)^{+}=L_{k_{1}}\dots L_{k_{n}},\qquad \left(a_{-n}\right)^{+}=a_{n},
\end{equation}
and the conjugation of the state $|P\rangle_{\vec{\lambda}}$ does not involve complex conjugation of its coefficients, i.e. for $|P\rangle_{\vec{\lambda}}$ given by \eqref{true-basis} we define conjugated state $_{\vec{\lambda}}\langle P|$ by
\begin{equation}
   _{\vec{\lambda}}\langle P|=\sum_{|\vec{\mu}|=|\vec{\lambda}|}C_{\vec{\lambda}}^{\mu_{1},\mu_{2}}(P)\,\langle P|
    (\hat{a}_{-\mu_{1}})^{+}(\hat{L}_{-\mu_{2}})^{+}.
\end{equation}
As it will be shown below, all the coefficients $C_{\vec{\lambda}}^{\mu_{1},\mu_{2}}(P)$  in \eqref{true-basis} are uniquely determined from the requirement that the matrix elements of primary field \eqref{primary} between any two states \eqref{true-basis} coincide with $Z_{\text{\sf bif}}$ defined above (equation \eqref{Zbif-def}).  
Here we arrive to the proposition which is the main result of our paper:
\begin{Proposition}\label{MAIN-Proposition}
There exists unique orthogonal basis $|P\rangle_{\vec{\lambda}}$ such that
\begin{equation}\label{matrix-elements}
    \frac{ _{\vec{\mu}}\langle P'|V_{\alpha}|P\rangle_{\vec{\lambda}}}
    {\langle P'|V_{\alpha}|P\rangle}=Z_{\text{\sf{bif}}}(\alpha|P',\vec{\mu};P,\vec{\lambda}).
\end{equation}
\end{Proposition}
The proof will be done in section \ref{Proof-section}. Here me make some comments which might be useful.
We note that equation \eqref{matrix-elements} with the r.h.s. given by \eqref{Zbif-def} can be considered as a system of equations for unknown coefficients $C_{\vec{\lambda}}^{\mu_{1},\mu_{2}}(P)$ in \eqref{true-basis}. Indeed inserting \eqref{true-basis} into \eqref{matrix-elements}
we obtain an overdetermined system. It is rather non-trivial that this system has a solution.  We found explicit form of the coefficients $C_{\vec{\lambda}}^{\mu_{1},\mu_{2}}(P)$ up to level $6$ (i.e.  for $|\vec{\lambda}|\leq6$). The first few representatives of the basis $|P\rangle_{\vec{\lambda}}$ obtained this way are:
\paragraph{level 1:}\label{Exp-examples}
\begin{equation*}
   \begin{aligned}
     &|P\rangle_{(1),\varnothing}=-\left(L_{-1}+i(Q+2P)a_{-1}\right)|P\rangle,\\
     &|P\rangle_{\varnothing,(1)}=-\left(L_{-1}+i(Q-2P)a_{-1}\right)|P\rangle,
   \end{aligned}
\end{equation*}
\paragraph{level 2:}
\begin{multline*}
     |P\rangle_{(2),\varnothing}=
     \left(L_{-1}^{2}-b^{-1}(Q+2P)L_{-2}+2i(Q+b^{-1}+2P)L_{-1}a_{-1}-\right.\\\left.-(Q + 2 P) (Q + b^{-1} + 2 P)a_{-1}^{2}
     -ib^{-1}(Q + 2 P) (Q + b^{-1} + 2 P)a_{-2}\right)|P\rangle,
\end{multline*}
\begin{multline*}
     |P\rangle_{\varnothing,(2)}=
     \left(L_{-1}^{2}-b^{-1}(Q-2P)L_{-2}+2i(Q+b^{-1}-2P)L_{-1}a_{-1}-\right.\\\left.-(Q-2 P) (Q+b^{-1}-2 P)a_{-1}^{2}
     -ib^{-1}(Q - 2 P) (Q + b^{-1} - 2 P)a_{-2}\right)|P\rangle,
\end{multline*}
\begin{multline*}
     |P\rangle_{(1,1),\varnothing}=
     \left(L_{-1}^{2}-b(Q+2P)L_{-2}+2i(Q+b+2P)L_{-1}a_{-1}-\right.\\\left.-(Q + 2 P) (Q+b+2P)a_{-1}^{2}
     -ib(Q + 2 P) (Q + b + 2 P)a_{-2}\right)|P\rangle,
\end{multline*}
\begin{multline*}
     |P\rangle_{\varnothing,(1,1)}=
     \left(L_{-1}^{2}-b(Q-2P)L_{-2}+2i(Q+b-2P)L_{-1}a_{-1}-\right.\\\left.-(Q - 2 P) (Q+b-2P)a_{-1}^{2}
     -ib(Q - 2 P) (Q + b - 2 P)a_{-2}\right)|P\rangle,
\end{multline*}
\begin{equation*}
     |P\rangle_{(1),(1)}=
     \left(L_{-1}^{2}-L_{-2}+2iQL_{-1}a_{-1}+(1+4P^{2}-Q^{2})a_{-1}^{2}-iQa_{-2}\right)|P\rangle.
\end{equation*}

We would like to mention here that it is rather  natural to expect that such an orthogonal basis could be a solution to the problem of simultaneous diagonalization of some infinite system of mutually commuting quantities (they are usually called Integrals of Motion). The r\^ole of IM's in conformal field theory was studied extensively  by Bazhanov, Lukyanov and Zamolodchikov in \cite{Bazhanov:1994ft,Bazhanov:1996dr,Bazhanov:1998dq}. In their case the system of quantum IM's was a ``quantization'' of KdV system. We were able to find an integrable system which corresponds to our case (with the algebra of symmetries being the tensor product of Virasoro and Heisenberg algebras). The results are collected in Appendix \ref{IM}. In particular, the classical counterpart of the quantum integrable system  is represented by \eqref{Classical-Integrable-Equation} which is known as $\text{Benjamin-Ono}_{\mathbf{\scriptscriptstyle{2}}}$ equation \cite{Lebedev-Radul,Degasperis,Degasperis2}.

Our strategy of proving of Proposition \ref{MAIN-Proposition} is the following. At first, we propose explicit expressions for  the states $|P\rangle_{\vec{\lambda}}$ in the case of second Young diagram being empty (i.e. $|P\rangle_{\lambda,\scriptscriptstyle{\varnothing}}$) and prove that the matrix elements between these proposed states have the form of the r.h.s. of \eqref{matrix-elements}. These states are expressed in terms of Jack polynomials. Then we define recursive procedure allowing to construct the rest of the basis i.e. all the states $|P\rangle_{\vec{\lambda}}$ with both general diagrams and also prove that the matrix elements are given by the r.h.s. of \eqref{matrix-elements}. And finally we prove the uniqueness of the basis $|P\rangle_{\vec{\lambda}}$. 
\section{Proof of Proposition \ref{MAIN-Proposition}}\label{Proof-section}
It would be very naive to expect to find  analytical expressions for all the states $|P\rangle_{\vec{\lambda}}$ in a closed form. However, the states of the form $|P\rangle_{\lambda,\scriptscriptstyle{\varnothing}}$ are particularly  simple\footnote{Similarly the states $|P\rangle_{\scriptscriptstyle{\varnothing},\lambda}$ are also simple (see below)}. They become even more simple if one expresses  the Virasoro generators $L_{n}$ in terms of bosons. Namely, let us represent Virasoro generators $L_{n}$ in terms of Heisenberg generators $c_{k}$  by
\begin{equation}\label{Bosonization}
  \begin{gathered}
   L_{n}=\sum_{k\neq0,n}c_{k}c_{n-k}+i(nQ-2\mathcal{P})c_{n},\quad L_{0}=\frac{Q^{2}}{4}-\mathcal{P}^{2}+2\sum_{k>0}c_{-k}c_{k},\\
    [c_{n},c_{m}]=\frac{n}{2}\,\delta_{n+m,0},\quad[\mathcal{P},c_{n}]=0,\quad\mathcal{P}|P\rangle=P|P\rangle,\quad\langle P|\mathcal{P}=-P\langle P|.
  \end{gathered}
\end{equation}
Based on explicit computations on lower levels we formulate the following proposition:
\begin{Proposition}\label{Jack-Proposition}
The  matrix elements between the states $|P\rangle_{\lambda,\scriptscriptstyle{\varnothing}}$ and $_{\mu, \scriptscriptstyle{\varnothing}}\langle P'|$ defined by
\begin{equation}\label{Jack-basis}
    |P\rangle_{\lambda,\scriptscriptstyle{\varnothing}}=\Omega_{\lambda}(P)\,\jac_{\lambda}^{\scriptscriptstyle{(1/g)}}(x)|P\rangle,\qquad
    _{\mu, \scriptscriptstyle{\varnothing}}\langle P'|=\Omega_{\mu}(P')\,\langle P'|\jac_{\mu}^{\scriptscriptstyle{(1/g)}}(y),
\end{equation}
are given by $Z_{\text{\sf bif}}(\alpha|P',(\mu,\varnothing);P,(\lambda,\varnothing))$. Here $g=-b^{2}$,
\begin{equation*}
     a_{-k}-c_{-k}=-ib\,p_{k}(x),\qquad
     a_{k}+c_{k}=-ib\,p_{k}(y),
\end{equation*}
with $p_{k}(x)$ being $k$-th power sum symmetric polynomial $p_{k}(x)=\sum_{j}x_{j}^{k}$ and $\jac_{\lambda}^{\scriptscriptstyle{(1/g)}}(x)$ is the Jack polynomial associated with the  Young diagram $\lambda$ normalized as (``integral form'' normalization \cite{Macdonald})
\begin{equation*}
   \jac_{\lambda}^{\scriptscriptstyle{(1/g)}}(x)=|\lambda|!\,m_{[1,\dots,1]}(x)+\dots,
\end{equation*}
where $m_{[\nu_{1},\dots,\nu_{n}]}(x)$ is the monomial symmetric polynomial.
\end{Proposition}
In \eqref{Jack-basis} the factor $\Omega_{\lambda}(P)$ is defined by
\begin{equation}\label{Omega}
   \Omega_{\lambda}(P)=(-b)^{|\lambda|}
    \prod_{(i,j)\in\lambda}(2P+ib+jb^{-1}),
\end{equation}
index $i$ runs vertically and $j$ runs horizontally over the diagram $\lambda$. For example, for the diagram $\lambda=(2,1)$ we have
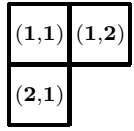
\begin{equation*}
  \begin{picture}(30,60)(60,20)
    \Thicklines
    \unitlength 2.3pt 
    \put(20,10){\line(0,1){20}}
    \put(20,10){\line(1,0){10}}
    \put(30,10){\line(0,1){20}}
    \put(20,20){\line(1,0){20}}
    \put(20,30){\line(1,0){20}}
    \put(40,20){\line(0,1){10}}
    \put(21,24){\mbox{$\scriptstyle{\mathbf{(1,1)}}$}}
    \put(31,24){\mbox{$\scriptstyle{\mathbf{(1,2)}}$}}
    \put(21,14){\mbox{$\scriptstyle{\mathbf{(2,1)}}$}}
    \end{picture}
\end{equation*}
Before proceed to  proof of Proposition \ref{Jack-Proposition}  let us make the following comment. In practice, the relation \eqref{Jack-basis} should be understood as follows. Take 
\begin{equation*}
    |P\rangle_{\lambda,\scriptscriptstyle{\varnothing}}=\sum_{|\mu_{1}|+|\mu_{2}|=|\lambda|}C_{\lambda}^{\mu_{1},\mu_{2}}(P)\,
     \hat{a}_{-\mu_{1}}\hat{L}_{-\mu_{2}}|P\rangle,
\end{equation*}
with some unknown coefficients $C_{\lambda}^{\mu_{1},\mu_{2}}(P)$. Then represent Virasoro generators $L_{k}$ in terms of bosons $c_{k}$ as in \eqref{Bosonization} and equate the result to the r.h.s. in \eqref{Jack-basis}. As a result all the coefficients $C_{\lambda}^{\mu_{1},\mu_{2}}(P)$ will be unambiguously determined provided that Verma module spanned by $\{L_{-\lambda}|P\rangle\}$ and Fock module spanned by $\{c_{-\lambda}|P\rangle\}$ are isomorphic for general values of the momenta $P$. Evidently, all the coefficients $C_{\lambda}^{\mu_{1},\mu_{2}}(P)$ are rational functions of the momenta $P$. In fact, due to the special choice of the factor $\Omega_{\lambda}(P)$ in \eqref{Jack-basis} all of them  are polynomials in $P$ (see for example explicit expressions for lower levels on the page \pageref{Exp-examples}). 

We note that we can choose different sign in front of operator $\mathcal{P}$ in \eqref{Bosonization}. This way we define another bosonization of the Virasoro algebra with generators $\hat{c}_{k}$. It is known that these two sets of generators $c_{k}$ and $\hat{c}_{k}$ are related by some unitary transform \cite{Zamolodchikov:1995aa}. We are not discussing this interesting question here but just mention that Proposition \ref{Jack-Proposition} is valid for the states $|P\rangle_{\scriptscriptstyle{\varnothing},\lambda}$ as well with $c_{k}\rightarrow \hat{c}_{k}$ and $\Omega_{\lambda}(P)\rightarrow \Omega_{\lambda}(-P)$. The proof is similar to what will be done below. We note that the states in Verma module of Virasoro algebra corresponding to Jack polynomials have been already studied in the literature \cite{Sakamoto:2004rn}.

\paragraph{Proof of Proposition \ref{Jack-Proposition}:}Let us show that the matrix elements between the states $|P\rangle_{\lambda,\scriptscriptstyle{\varnothing}}$ defined by \eqref{Jack-basis} 
\begin{equation}\label{needed-matrix-element}
   \frac{ _{\mu,\scriptscriptstyle{\varnothing}}\langle P'|V_{\alpha}|P\rangle_{\lambda,\scriptscriptstyle{\varnothing}}}
    {\langle P'|V_{\alpha}|P\rangle}.
\end{equation}
are given by  \eqref{matrix-elements}. We note that as follows from \eqref{commutation-relations} and \eqref{form-of-3point-function} the matrix element \eqref{needed-matrix-element} is a polynomial in $\alpha$ of degree $2|\lambda|+2|\mu|$. Henceforth, in order to restore this polynomial it is enough to find its values at  $(2|\lambda|+2|\mu|+1)$ distinct points. In fact, using representation \eqref{Jack-basis} we can find it at infinite set of points $\alpha_{n}$ which solve the following screening condition
\begin{equation}\label{Scr-condition}
    P+P'+\alpha+nb=0,\quad\text{with}\quad n\in\mathbb{Z}_{\geq0}.
\end{equation}
In this case the matrix element \eqref{needed-matrix-element} possesses the free-field representation \cite{Kanie-Tsuchiya,Dotsenko:1984nm,Felder:1988zp}. Namely, one have to introduce the screening charge
\begin{equation}
    \mathcal{S}=\oint\limits_{\mathcal{C}}e^{2b\Phi(\xi)}d\xi,\qquad \Phi(\xi)=-i\mathcal{P}\log(\xi)+i\sum_{k\neq0}\frac{c_{k}}{k}\,\xi^{-k},
\end{equation}
which commutes with the Virasoro algebra and define the screened vertex operator
\begin{equation}\label{vertex-definition}
   V_{\alpha_{n}}^{L}(z)\longrightarrow \mathcal{S}^{n}e^{2\alpha_{n}\Phi(z)},
\end{equation}
where the contours of integrations are started from the point $z$ and go around $0$ counterclockwise. Representing the primary field $V_{\alpha_{n}}=\mathcal{V}_{\alpha_{n}}\cdot V_{\alpha_{n}}^{L}$ in terms of free fields we can proceed and compute the matrix element \eqref{needed-matrix-element} using the commutation relations of two Heisenberg algebras (with $a_{k}$ and $c_{k}$ generators). We note that computation simplifies drastically since the operator creating the ket state $|P\rangle_{\lambda,\scriptscriptstyle{\varnothing}}$ commutes with the operator creating the bra state $_{\mu,\scriptscriptstyle{\varnothing}}\langle P|$ as one of them depends on difference of bosons $a_{-k}-c_{-k}$ while another on sum $a_{k}+c_{k}$. After completing the algebraic part of this exercise (we skip the details here due to their triviality) we are left with the problem of computation of some multiple contour integral\footnote{We note that inside correlation functions the  contours can be deformed to those considered in \cite{Dotsenko:1984ad,Dotsenko:1984nm}. This deformation of the contours gives some factors arising from the non-analyticity of the integrand. In our case these factors are not important since we consider the ratio of the three-point correlation function involving descendant fields and the three-point function of primary fields (see \eqref{Generalized-Selberg} below). Evidently, they cancel each other in the ratio.}. It can be summarized as follows.
Let $\alpha_{n}$ solves the equation \eqref{Scr-condition} then the matrix elements between the states $|P\rangle_{\lambda,\scriptscriptstyle{\varnothing}}$ and $_{\mu, \scriptscriptstyle{\varnothing}}\langle P'|$ defined by \eqref{Jack-basis} can be written as
\begin{equation}\label{Generalized-Selberg}
    \frac{ _{\mu, \scriptscriptstyle{\varnothing}}\langle P'|V_{\alpha_{n}}|P\rangle_{\lambda,\scriptscriptstyle{\varnothing}}}
    {\langle P'|V_{\alpha_{n}}|P\rangle}=\Omega_{\lambda}(P)\Omega_{\mu}(P')\,
    \frac{\left\langle\jac_{\mu}^{\scriptscriptstyle{(1/g)}}[p_{k}+\rho]\,\jac_{\lambda}^{\scriptscriptstyle{(1/g)}}[p_{-k}]\right\rangle_{\text{\sf Sel}}^{(n)}}
    {\langle1\rangle_{\text{\sf Sel}}^{(n)}},  
\end{equation}
where $\rho=(2\alpha_{n}-Q)/b$ and  $\langle\dots\rangle_{\text{\sf Sel}}^{(n)}$ denotes the  Selberg average
\begin{equation*}
   \langle \mathcal{O}\rangle_{\text{\sf Sel}}^{(n)}\overset{\text{def}}{=}\frac{1}{n!}
   \int_{0}^{1}\dots\int_{0}^{1}\, \mathcal{O}(t_{1},\dots,t_{n})\,\prod_{j=1}^{n}t_{j}^{A}(1-t_{j})^{B}\prod_{i<j}|t_{i}-t_{j}|^{2g} dt_{1}\dots dt_{n},
\end{equation*}
with parameters $A$, $B$ and $g$ given by
\begin{equation}\label{ABg}
  A=-b(Q+2P),\qquad B=-2b\alpha_{n}\quad\text{and}\quad g=-b^{2}.
\end{equation}
In the r.h.s. in \eqref{Generalized-Selberg} we use the following notation from the symmetric functions theory\label{Plethystic}\footnote{The notation $f[s_{k}]$ (square brackets instead of parentheses)  is known in symmetric functions theory as plethystic substitution \cite{Rains}. The more common notation used in the literature is $f[s]$ (see monographs \cite{Haglund,Lascoux}).  We thank referees for suggesting to comment on this notation.}: for any symmetric function $f(x_{1},\dots,x_{n})$ we define the notation $f[s_{k}]$ which means that the evaluation homomorphism which sends $k$-th power sum symmetric polynomial $p_{k}$ to $s_{k}$ has been applied to $f$.

As it will be shown in appendix \ref{Selberg-proof}, the integral in the r.h.s. in \eqref{Generalized-Selberg} can be computed exactly with the expected result
\begin{equation}\label{Selberg-matrix-element}
       \frac{ _{\mu, \scriptscriptstyle{\varnothing}}\langle P'|V_{\alpha_{n}}|P\rangle_{\lambda,\scriptscriptstyle{\varnothing}}}
    {\langle P'|V_{\alpha_{n}}|P\rangle} =Z_{\text{\sf{bif}}}(\alpha_{n}|P',(\mu,\varnothing);P,(\lambda,\varnothing)),
\end{equation}
where the function $Z_{\text{\sf{bif}}}(\alpha|P',(\mu,\varnothing);P,(\lambda,\varnothing))$ is defined by \eqref{Zbif-def}.
As was explained above  we can continue  \eqref{Selberg-matrix-element} to arbitrary values of $\alpha$ since the matrix element \eqref{needed-matrix-element} is a polynomial in $\alpha$. It proves that the matrix elements between the states $|P\rangle_{\lambda, \scriptscriptstyle{\varnothing}}$ and $_{\mu,\scriptscriptstyle{\varnothing}}\langle P'|$ defined in Proposition \ref{Jack-Proposition} have desired factorized form \eqref{matrix-elements} for arbitrary values of the parameters $P$, $P'$ and $\alpha$. $\square$

So far we have constructed only the states $|P\rangle_{\vec{\lambda}}$ for the pairs of Young diagrams of the form $(\lambda,\varnothing)$. Now we are going to define the recursive procedure allowing to construct the rest of the basis.  We note that the state $|P\rangle_{\lambda,\scriptscriptstyle{\varnothing}}$ expressed in terms of Heisenberg generators $a_{k}$ and $c_{k}$ as in \eqref{Jack-basis} vanishes due to the factor \eqref{Omega} for 
\begin{equation}
    P=P_{m,n}\overset{\text{def}}{=}-\frac{mb+nb^{-1}}{2},\qquad\text{for}\quad (m,n)\in \lambda,
\end{equation}
i.e. at the value of the momenta $P$ such that corresponding Verma module become degenerate \cite{Belavin:1984vu}. Namely, for $P=P_{m,n}$ there exists a singular vector $|\chi_{m,n}\rangle$ in Verma module $|P_{m,n}\rangle$ at the level $mn$
\begin{equation}
    |\chi_{m,n}\rangle\overset{\text{def}}{=}D_{m,n}|P_{m,n}\rangle=\left(L_{-1}^{mn}+\dots\right)|P_{m,n}\rangle,
\end{equation}
such that $L_{k}|\chi_{m,n}\rangle=0$ for any $k>0$.
However, the state $|P\rangle_{\lambda,\scriptscriptstyle{\varnothing}}$ does not vanish when expressed in terms of generators $L_{n}$ and $a_{n}$ instead of $c_{n}$ and $a_{n}$. Indeed we have proved that the matrix elements between the states $|P\rangle_{\lambda,\scriptscriptstyle{\varnothing}}$ are given by \eqref{matrix-elements}. In particular,
\begin{equation}\label{Jack-simple-matrix-element}
  \frac{\langle P'|V_{\alpha}|P\rangle_{\lambda,\scriptscriptstyle{\varnothing}}}
    {\langle P'|V_{\alpha}|P\rangle} =Z_{\text{\sf{bif}}}(\alpha|P',(\varnothing,\varnothing);P,(\lambda,\varnothing)).
\end{equation}
Comparing the behavior of both hand sides of \eqref{Jack-simple-matrix-element} at $\alpha\rightarrow\infty$ one can estimate the coefficient in $|P\rangle_{\lambda,\scriptscriptstyle{\varnothing}}$ in front of $L_{-1}^{|\lambda|}$. Using \eqref{Zbif-def} and \eqref{commutation-relations} we find 
\begin{equation*}
   |P\rangle_{\lambda,\scriptscriptstyle{\varnothing}}=((-L_{-1})^{|\lambda|}+\dots)|P\rangle,
\end{equation*}
where omitted terms have degree in $L_{-1}$ at most $(|\lambda|-1)$. We  see that the coefficient before $L_{-1}^{|\lambda|}$ does not vanish for any values of $P$ and hence the state $|P\rangle_{\lambda,\scriptscriptstyle{\varnothing}}$ as well\footnote{One can not exclude the possibility that the state $|P\rangle_{\lambda,\scriptscriptstyle{\varnothing}}$ has a pole at some $P=P_{m,n}$. We will see below that this is not the case.}. It was shown by Feigin and Fuks \cite{Feigin-Fuks} that a state which does not vanish in Verma module, but vanishes after bosonization  in Fock module  is some descendant of the singular vector $|\chi_{m,n}\rangle$. We arrive to the following proposition:
\begin{Proposition}\label{X-prop}
Let $\lambda=(\lambda_{1},\lambda_{2},\dots)$ a partition and  $|P\rangle_{\lambda,\scriptscriptstyle{\varnothing}}$ the state defined by \eqref{Jack-basis}  then the state $|P=P_{m,n}\rangle_{\lambda,\scriptscriptstyle{\varnothing}}$ for $(m,n)\in\lambda$ has a factorized form
\begin{equation}\label{XX-def}
   |P_{m,n}\rangle_{\lambda,\scriptscriptstyle{\varnothing}}=(-1)^{mn}X_{\lambda}^{(m,n)}\,D_{m,n}|P_{m,n}\rangle
   \quad\text{for}\quad (m,n)\in\lambda,
\end{equation}
where
\begin{equation*}
X_{\lambda}^{(m,n)}=\sum_{|\vec{\sigma}|=|\lambda|-mn}C^{(m,n)}_{\lambda;\,\vec{\sigma}}\,
    \hat{a}_{-\sigma_{1}}\hat{L}_{-\sigma_{2}},
\end{equation*}   
is the operator which satisfies
\begin{equation}\label{X-prop-statement}
     \frac{_{\mu,\scriptscriptstyle{\varnothing}}\langle P'|V_{\alpha}\,X_{\lambda}^{(m,n)}|P_{m,-n}\rangle}{\langle P'|V_{\alpha}|P_{m,-n}\rangle}=
     Z_{\text{\sf bif}}(\alpha|P',(\mu,\varnothing);P_{m,-n},(\rho,\nu)),
\end{equation}
and the pair of partitions $(\rho,\nu)$  is defined as $\rho=(\lambda_{1}-n,\dots,\lambda_{m}-n)$ and $\nu=(\lambda_{m+1},\lambda_{m+2},\dots)$.
\end{Proposition}
An example of how the pair of the partitions $(\rho,\nu)$ is defined for given $(m,n)\in\lambda$ is shown by the following picture.
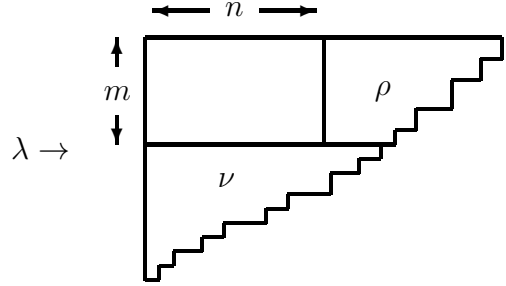
\begin{equation*}
  \begin{picture}(50,120)(90,-90)
    \Thicklines
    \unitlength 2.7pt 
   \rotatebox{270}{
    \put(0,20){\line(0,1){50}}
    \put(15,20){\line(0,1){35}}
    \put(0,20){\line(1,0){34}}
    \put(0,45){\line(1,0){15}}
    \put(0,70){\line(1,0){3}}
    \put(3,70){\line(0,-1){3}}
    \put(3,67){\line(1,0){3}}
    \put(6,67){\line(0,-1){4}}
    \put(6,63){\line(1,0){4}}
    \put(10,63){\line(0,-1){5}}
    \put(10,58){\line(1,0){3}}
    \put(13,58){\line(0,-1){3}}
    \put(13,55){\line(1,0){2}}
    \put(15,53){\line(1,0){2}}
    \put(17,53){\line(0,-1){3}}
    \put(17,50){\line(1,0){2}}
    \put(19,50){\line(0,-1){4}}
    \put(19,46){\line(1,0){3}}
    \put(22,46){\line(0,-1){6}}
    \put(22,40){\line(1,0){2}}
    \put(24,40){\line(0,-1){3}}
    \put(24,37){\line(1,0){2}}
    \put(26,37){\line(0,-1){6}}
    \put(26,31){\line(1,0){2}}
    \put(28,31){\line(0,-1){3}}
    \put(28,28){\line(1,0){2}}
    \put(30,28){\line(0,-1){4}}
    \put(30,24){\line(1,0){2}}
    \put(32,24){\line(0,-1){2}}
    \put(32,22){\line(1,0){2}}
    \put(34,22){\line(0,-1){2}}
    \put(11,16){\vector(1,0){4}}
    \put(4,16){\vector(-1,0){4}}
    \put(-3.8,36){\vector(0,1){8}}
    \put(-3.8,29){\vector(0,-1){8}}
    }
    \put(-18,-8){\mbox{$\rho$}}
    \put(-40,-21){\mbox{$\nu$}}
    \put(-56,-8.8){\mbox{$m$}}
    \put(-39,3){\mbox{$n$}}
    \put(-69,-17){\mbox{$\lambda\rightarrow$}}
    \end{picture}
\end{equation*}
\paragraph{Proof of Proposition \ref{X-prop}:}As was explained above \eqref{XX-def} follows from the results of Feigin and Fuks \cite{Feigin-Fuks}. In order to prove \eqref{X-prop-statement} we start with
\begin{equation}\label{Selberg-matrix-element-X-prop}
       \frac{ _{\mu, \scriptscriptstyle{\varnothing}}\langle P'|V_{\alpha}|P\rangle_{\lambda,\scriptscriptstyle{\varnothing}}}
    {\langle P'|V_{\alpha}|P\rangle} =Z_{\text{\sf{bif}}}(\alpha|P',(\mu,\varnothing);P,(\lambda,\varnothing)),
\end{equation}
Plugging $P=P_{m,n}$ for $(m,n)\in\lambda$ and using \eqref{XX-def} we find
\begin{multline}\label{matrix-element-degenerate}
     \frac{ _{\mu, \scriptscriptstyle{\varnothing}}\langle P'|V_{\alpha}|P_{m,n}\rangle_{\lambda,\scriptscriptstyle{\varnothing}}}
    {\langle P'|V_{\alpha}|P_{m,n}\rangle}\overset{\eqref{XX-def}}{=}(-1)^{mn}
    \frac{ _{\mu, \scriptscriptstyle{\varnothing}}\langle P'|V_{\alpha}\,X_{\lambda}^{(m,n)}D_{m,n}|P_{m,n}\rangle}
    {\langle P'|V_{\alpha}|P_{m,n}\rangle}=\\
    =(-1)^{mn}
     \frac{_{\mu,\scriptscriptstyle{\varnothing}}\langle P'|V_{\alpha}\,X_{\lambda}^{(m,n)}|P_{m,-n}\rangle}{\langle P'|V_{\alpha}|P_{m,-n}\rangle}
     \frac{\langle P'|V_{\alpha}D_{m,n}|P_{m,n}\rangle}{\langle P'|V_{\alpha}|P_{m,n}\rangle}.
\end{multline}
In the second line in \eqref{matrix-element-degenerate} we used  the equality $\Delta(P_{m,-n})=\Delta(P_{m,n})+mn$ and the fact that $L_{k}D_{m,n}|P_{m,n}\rangle=a_{k}D_{m,n}|P_{m,n}\rangle=0$ for $k>0$ provided that $D_{m,n}|P_{m,n}\rangle$ is a singular vector.
The last factor in the second line in \eqref{matrix-element-degenerate} is known explicitly \cite{Zamolodchikov:2005fy}
\begin{equation}\label{Pmn-polynomial}
   \frac{\langle P'|V_{\alpha}\,D_{m,n}|P_{m,n}\rangle}{\langle P'|V_{\alpha}|P_{m,n}\rangle}\overset{\text{def}}{=}
   \mathbb{P}_{m,n}(\alpha,P')=
   p_{m,n}\left(\frac{Q}{2}+P'-\alpha\right)p_{m,n}\left(\alpha+P'-\frac{Q}{2}\right),
\end{equation}
where
\begin{equation*}
   p_{m,n}(x)=\prod_{r,s}(x+P_{r,s}),
\end{equation*}
and the pair of integers $r,s$ runs over the set 
\begin{equation*}
    \begin{aligned}
       &r=-m+1,-m+3,\dots,m-3,m-1,\\
       &s=-n+1,-n+3,\dots,n-3,n-1.
    \end{aligned}
\end{equation*}
Using \eqref{Selberg-matrix-element-X-prop}, \eqref{matrix-element-degenerate} and  the simple identity (see short proof in Appendix \ref{ME-reduction})
\begin{equation}\label{Z-Z-relat}
    Z_{\text{\sf bif}}(\alpha|P',(\mu,\varnothing);P_{m,n},(\lambda,\varnothing))=(-1)^{mn}\,\mathbb{P}_{m,n}(\alpha,P')\,
    Z_{\text{\sf bif}}(\alpha|P',(\mu,\varnothing);P_{m,-n},(\rho,\nu)),
\end{equation}
with $\rho=(\lambda_{1}-n,\dots,\lambda_{m}-n)$ and $\nu=(\lambda_{m+1},\lambda_{m+2},\dots)$    we find
\begin{equation}\label{X-prop-statement-2}
     \frac{_{\mu,\scriptscriptstyle{\varnothing}}\langle P'|V_{\alpha}\,X_{\lambda}^{(m,n)}|P_{m,-n}\rangle}{\langle P'|V_{\alpha}|P_{m,-n}\rangle}=
     Z_{\text{\sf bif}}(\alpha|P',(\mu,\varnothing);P_{m,-n},(\rho,\nu)),
\end{equation}
thus completing the proof of Proposition \ref{X-prop}. $\square$
\begin{Proposition}\label{X-General-prop}
For any pair of Young diagrams $(\rho,\nu)$ there exists unique operator
\begin{equation}\label{X-General}
    \mathbf{X}_{\rho,\nu}(P)=\sum_{|\vec{\sigma}|=|\rho|+|\nu|}C_{\rho,\nu}^{\vec{\sigma}}(P)\,
    \hat{a}_{-\sigma_{1}}\hat{L}_{-\sigma_{2}},
\end{equation}
such that
\begin{subequations}
\begin{equation}\label{Jack-General-matrix-element}
     \frac{_{\mu,\scriptscriptstyle{\varnothing}}\langle P'|V_{\alpha}\,\mathbf{X}_{\rho,\nu}(P)|P\rangle}{\langle P'|V_{\alpha}|P\rangle}=
     Z_{\text{\sf bif}}(\alpha|P',(\mu,\varnothing);P,(\rho,\nu)),\quad\forall\;\mu
\end{equation}
where $_{\mu,\scriptscriptstyle{\varnothing}}\langle P'|$ is defined by \eqref{Jack-basis}.  Moreover, for $m\geq l(\rho)$, $n\geq \nu_{1}-\rho_{m}$
\begin{equation}\label{Jack-General-matrix-element-2}
    \mathbf{X}_{\rho,\nu}(P_{m,-n})=X_{\lambda}^{(m,n)},
\end{equation}
where $\lambda=(\rho_{1}+n,\dots,\rho_{m}+n,\nu_{1},\nu_{2},\dots)$ and $X_{\lambda}^{(m,n)}$ is the operator defined in Proposition \ref{X-prop}.
\end{subequations}
\end{Proposition}
\paragraph{Proof of Proposition \ref{X-General-prop}:}The existence of the  operator $\mathbf{X}_{\rho,\nu}(P)$ is equivalent to the statement that the infinite system of linear equations \eqref{Jack-General-matrix-element} on coefficients $C_{\rho,\nu}^{\vec{\sigma}}(P)$  has a solution. Indeed \eqref{Jack-General-matrix-element} can be written as
\begin{equation}\label{Linear-system-Jack}
   \sum_{|\vec{\sigma}|=|\rho|+|\nu|}C_{\rho,\nu}^{\vec{\sigma}}(P)\mathcal{M}_{\vec{\sigma},\mu}(\alpha,P,P')=
   Z_{\text{\sf bif}}(\alpha|P',(\mu,\varnothing);P,(\rho,\nu)),
\end{equation}
where 
\begin{equation*}
    \mathcal{M}_{\vec{\sigma},\mu}(\alpha,P,P')=
    \frac{_{\mu,\scriptscriptstyle{\varnothing}}\langle P'|V_{\alpha}\,\hat{a}_{-\sigma_{1}}\hat{L}_{-\sigma_{2}}|P\rangle}{\langle P'|V_{\alpha}|P\rangle}.
\end{equation*}
The equation \eqref{Linear-system-Jack} should be valid for any partition $\mu$. One can rewrite \eqref{Linear-system-Jack} in a matrix form and  simplify it by Gaussian elimination method reducing it to either row echelon form (having infinite number of zero rows), or to some degenerate system with no solutions. In any case the resulting system will be with coefficients which are rational functions of the momenta $P$. As was claimed in Proposition \ref{X-prop} this system has a solution at $P=P_{m,-n}$ for an \emph{infinite} set of integers $m$ and $n$.  Since the coefficients in \eqref{Linear-system-Jack} and hence in the reduced system are rational functions of $P$ this implies that \eqref{Linear-system-Jack} has solutions for general values of $P$. 

In fact, the reduced system has maximal rank and hence the system \eqref{Linear-system-Jack} has a unique solution. In order to prove this fact let us suppose that there are two operators $\mathbf{X}_{\rho,\nu}(P)$ and $\tilde{\mathbf{X}}_{\rho,\nu}(P)$ satisfying \eqref{Jack-General-matrix-element} and take $Y_{N}(P)=\mathbf{X}_{\rho,\nu}(P)-\tilde{\mathbf{X}}_{\rho,\nu}(P)$ where $N=|\rho|+|\nu|$.  Then \eqref{Jack-General-matrix-element} implies
\begin{equation}\label{X-N-null-cond}
  _{\mu,\scriptscriptstyle{\varnothing}}\langle P'|V_{\alpha}\,Y_{N}(P)|P\rangle=0,\quad\forall\;\mu.
\end{equation}
Consider Young diagrams $\mu$ of special form
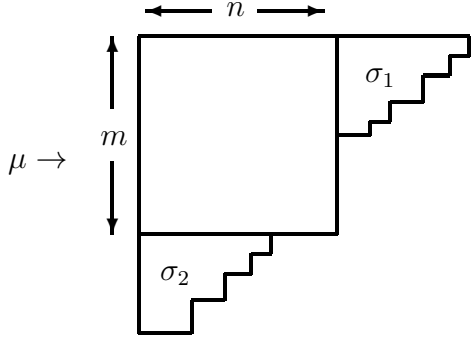
\begin{equation*}
    \begin{picture}(50,140)(60,-120)
    \Thicklines
    \unitlength 2.5pt 
    \rotatebox{270}{
    \put(0,50){\line(0,1){20}}
    \put(0,50){\line(1,0){15}}
    \put(15,50){\line(0,1){5}}
    \put(0,70){\line(1,0){3}}
    \put(3,70){\line(0,-1){3}}
    \put(3,67){\line(1,0){3}}
    \put(6,67){\line(0,-1){4}}
    \put(6,63){\line(1,0){4}}
    \put(10,63){\line(0,-1){5}}
    \put(10,58){\line(1,0){3}}
    \put(13,58){\line(0,-1){3}}
    \put(13,55){\line(1,0){2}}
    \put(30,20){\line(0,1){20}}
    \put(30,20){\line(1,0){15}}
    \put(45,20){\line(0,1){5}}
    \put(30,40){\line(1,0){3}}
    \put(33,40){\line(0,-1){3}}
    \put(33,37){\line(1,0){3}}
    \put(36,37){\line(0,-1){4}}
    \put(36,33){\line(1,0){4}}
    \put(40,33){\line(0,-1){5}}
    \put(40,28){\line(1,0){5}}
    \put(45,28){\line(0,-1){3}}
    \put(0,20){\line(0,1){40}}
    \put(0,20){\line(1,0){40}}
    \put(30,20){\line(0,1){30}}
    \put(30,50){\line(-1,0){30}}
    \put(19,16){\vector(1,0){11}}
    \put(11,16){\vector(-1,0){11}}
    \put(-3.8,38){\vector(0,1){10}}
    \put(-3.8,31){\vector(0,-1){10}}
    }
    \put(-16,-7){\mbox{$\sigma_{1}$}}
    \put(-47,-37){\mbox{$\sigma_{2}$}}
    \put(-56,-16.5){\mbox{$m$}}
    \put(-37,3){\mbox{$n$}}
    \put(-70,-20){\mbox{$\mu\rightarrow$}}
    \end{picture}
\end{equation*}
with arbitrary integers $m$ and $n$ such that $m,n\geq N$ and $|\vec{\sigma}|<N$. Due to Proposition \ref{X-prop} we have
\begin{equation}\label{XX-def-dual}
   _{\mu,\scriptscriptstyle{\varnothing}}\langle P_{m,n}|=(-1)^{mn}\langle P_{m,n}|D_{m,n}^{+}(X_{\mu}^{(m,n)})^{+},
\end{equation}
then \eqref{X-N-null-cond} and \eqref{XX-def-dual} imply
\begin{equation}\label{X-N-null-cond-2}
   \langle P_{m,-n}|(X_{\mu}^{(m,n)})^{+}\,V_{\alpha}\,Y_{N}(P)|P\rangle=0.
\end{equation}
We stress that for given value of $(|\mu|-mn)$ the states $\langle P_{m,-n}|(X_{\mu}^{(m,n)})^{+}$ are linearly independent provided that the Jack polynomials $\mathbf{J}^{\scriptscriptstyle{(1/g)}}_{\mu}(x)$ are independent for different partitions $\mu$. Moreover as one can readily see from the picture above  the number of states $\langle P_{m,-n}|(X_{\mu}^{(m,n)})^{+}$ with $|\mu|-mn=k<N$ equals to the number of pairs of Young diagrams $\vec{\sigma}$ with $|\vec{\sigma}|=k$, i.e. they form a basis. In other words any state $\langle P_{m,-n}|(X_{\mu}^{(m,n)})^{+}$ with $|\mu|-mn=k<N$ can be written as linear combination of the states $\langle P_{m,-n}|\hat{L}_{\mu_{1}}\hat{a}_{\mu_{2}}$ with $|\mu_{1}|+|\mu_{2}|=k$. Henceforth, equation \eqref{X-N-null-cond-2} is equivalent to
\begin{equation}\label{X-N-null-cond-3}
   \langle P_{m,-n}|\hat{L}_{\mu_{1}}\hat{a}_{\mu_{2}}\,V_{\alpha}\,Y_{N}(P)|P\rangle=0\qquad\text{for any}\;\mu_{1}, \mu_{2}:
    \;\;|\mu_{1}|+|\mu_{2}|< N.
\end{equation}
The state $Y_{N}(P)|P\rangle$ which satisfies \eqref{X-N-null-cond-3} is identically zero due to the following lemma: 
\begin{lemma}\label{lemma}
Let $Y_{N}(P)$ satisfy
\begin{equation}\label{X-N-null-cond-lemma}
   \langle P'|\hat{L}_{\mu_{1}}\hat{a}_{\mu_{2}}\,V_{\alpha}\,Y_{N}(P)|P\rangle=0\qquad\text{for any}\;\;\mu_{1}, \mu_{2}:
    \;\;|\mu_{1}|+|\mu_{2}|< N,
\end{equation}
and any parameters $P'$ and $\alpha$ then $Y_{N}(P)=0$.
\end{lemma}
The proof is based on induction. For $N=1,2$ this statement can be proved by straightforward computation. Assume that lemma holds  for $n=1,\dots,N-1$ and take
\begin{equation}\label{X-null-condition-3}
    \langle P'|\hat{L}_{\mu_{1}}\hat{a}_{\mu_{2}}a_{k}\,V_{\alpha}\,Y_{N}(P)|P\rangle\qquad\text{for any}\;\;\mu_{1}, \mu_{2}:
    \;\;|\mu_{1}|+|\mu_{2}|< N-k,\quad k=1,\dots,N.
\end{equation}
The matrix element in \eqref{X-null-condition-3} is equal to zero due to lemma assumption \eqref{X-N-null-cond-lemma}. From other side
\begin{multline}\label{X-null-condition-4}
   \langle P'|\hat{L}_{\mu_{1}}\hat{a}_{\mu_{2}}a_{k}\,V_{\alpha}\,Y_{N}(P)|P\rangle=
   \langle P'|\hat{L}_{\mu_{1}}\hat{a}_{\mu_{2}}[a_{k},\,V_{\alpha}]\,Y_{N}(P)|P\rangle+\langle P'|\hat{L}_{\mu_{1}}\hat{a}_{\mu_{2}}\,V_{\alpha}\,a_{k}\,Y_{N}(P)|P\rangle\overset{\eqref{commutation-relations}}{=}\\\overset{\eqref{commutation-relations}}{=}
   i(Q-\alpha)\langle P'|\hat{L}_{\mu_{1}}\hat{a}_{\mu_{2}}\,V_{\alpha}\,Y_{N}(P)|P\rangle+\langle P'|\hat{L}_{\mu_{1}}\hat{a}_{\mu_{2}}\,V_{\alpha}\,a_{k}\,Y_{N}(P)|P\rangle
\end{multline}
First term in the second line in \eqref{X-null-condition-4} is zero due to \eqref{X-N-null-cond-lemma}, which implies
\begin{equation}
  \langle P'|\hat{L}_{\mu_{1}}\hat{a}_{\mu_{2}}\,V_{\alpha}\,a_{k}\,Y_{N}(P)|P\rangle=0\qquad\text{for any}\;\mu_{1}, \mu_{2}:
    \;\;|\mu_{1}|+|\mu_{2}|\leq N-k,\quad k=1,\dots,N,
\end{equation}
and hence under induction assumptions $a_{k}\,Y_{N}(P)|P\rangle=0$ for $k=1,\dots,N-1$. Equivalently, we show that $L_{k}\,Y_{N}(P)|P\rangle=0$ for $k=1,\dots,N-1$ as well. Henceforth, we conclude that for general value of the momenta $P$ the only possibility left is  $Y_{N}(P)=Ca_{-N}$. Taking \eqref{X-N-null-cond-lemma} for $\mu_{1}=\mu_{2}=\varnothing$ and using \eqref{commutation-relations} we find that $C=0$. $\square$

It follows from  Lemma \ref{lemma} from that the system \eqref{Linear-system-Jack} has a unique solution thus the proof of Proposition \ref{X-General-prop} is complete. $\square$ We note that the fact that the system \eqref{Linear-system-Jack} has a unique solution implies that the coefficients $C_{\rho,\nu}^{\vec{\sigma}}(P)$ in \eqref{X-General} are rational functions of the momenta $P$. It is worthwhile to mention that Proposition \ref{X-General-prop} implies that
\begin{equation*}
    \mathbf{X}_{\lambda,\scriptscriptstyle{\varnothing}}|P\rangle=|P\rangle_{\lambda,\scriptscriptstyle{\varnothing}},
\end{equation*}
where $|P\rangle_{\lambda,\scriptscriptstyle{\varnothing}}$ is the state defined in \eqref{Jack-basis}.
\begin{Proposition}\label{XX-prop}
For any two pairs of Young diagrams $\vec{\lambda}=(\lambda_{1},\lambda_{2})$ and $\vec{\mu}=(\mu_{1},\mu_{2})$
\begin{equation}\label{General-General-matrix-element}
     \frac{\langle P'|\mathbf{X}^{\scriptscriptstyle{+}}_{\vec{\mu}}(P')\,V_{\alpha}\,\mathbf{X}_{\vec{\lambda}}(P)|P\rangle}{\langle P'|V_{\alpha}|P\rangle}=
     Z_{\text{\sf bif}}(\alpha|P',\vec{\mu};P,\vec{\lambda}).
\end{equation}
\end{Proposition}
\paragraph{Proof of Proposition \ref{XX-prop}:}In order to prove Proposition \ref{XX-prop} it is enough to show that \eqref{General-General-matrix-element} is valid for $P'=P_{m,-n}$ for some infinite set of integers $m$ and $n$ provided that the the coefficient $C_{\vec{\lambda}}^{\vec{\nu}}(P)$ in \eqref{X-General} are rational functions of the momenta $P$. For $\vec{\mu}=(\mu_{1},\mu_{2})$ take a ``master'' partition $\nu$ as
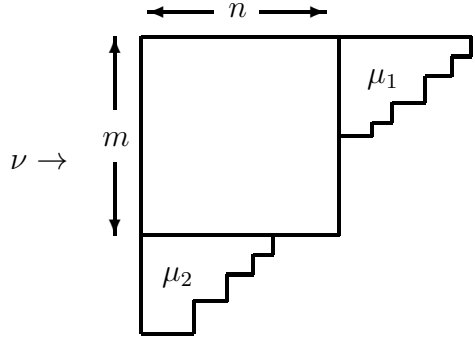
\begin{equation*}
    \begin{picture}(50,140)(60,-120)
    \Thicklines
    \unitlength 2.5pt 
    \rotatebox{270}{
    \put(0,50){\line(0,1){20}}
    \put(0,50){\line(1,0){15}}
    \put(15,50){\line(0,1){5}}
    \put(0,70){\line(1,0){3}}
    \put(3,70){\line(0,-1){3}}
    \put(3,67){\line(1,0){3}}
    \put(6,67){\line(0,-1){4}}
    \put(6,63){\line(1,0){4}}
    \put(10,63){\line(0,-1){5}}
    \put(10,58){\line(1,0){3}}
    \put(13,58){\line(0,-1){3}}
    \put(13,55){\line(1,0){2}}
    \put(30,20){\line(0,1){20}}
    \put(30,20){\line(1,0){15}}
    \put(45,20){\line(0,1){5}}
    \put(30,40){\line(1,0){3}}
    \put(33,40){\line(0,-1){3}}
    \put(33,37){\line(1,0){3}}
    \put(36,37){\line(0,-1){4}}
    \put(36,33){\line(1,0){4}}
    \put(40,33){\line(0,-1){5}}
    \put(40,28){\line(1,0){5}}
    \put(45,28){\line(0,-1){3}}
    \put(0,20){\line(0,1){40}}
    \put(0,20){\line(1,0){40}}
    \put(30,20){\line(0,1){30}}
    \put(30,50){\line(-1,0){30}}
    \put(19,16){\vector(1,0){11}}
    \put(11,16){\vector(-1,0){11}}
    \put(-3.8,38){\vector(0,1){10}}
    \put(-3.8,31){\vector(0,-1){10}}
    }
    \put(-16,-7){\mbox{$\mu_{1}$}}
    \put(-47,-37){\mbox{$\mu_{2}$}}
    \put(-56,-16.5){\mbox{$m$}}
    \put(-37,3){\mbox{$n$}}
    \put(-70,-20){\mbox{$\nu\rightarrow$}}
    \end{picture}
\end{equation*}
where $m$ and $n$ are large enough to ``fit'' the partitions $\mu_{1}$ and $\mu_{2}$. Then using conjugated version of \eqref{Jack-General-matrix-element-2}
\begin{equation}
  \mathbf{X}^{\scriptscriptstyle{+}}_{\vec{\mu}}(P_{m,-n})=(X^{(m,n)}_{\nu})^{\scriptscriptstyle{+}},
\end{equation}
we find
\begin{multline}\label{XX-prop-proof-1}
   \frac{\langle P_{m,-n}|\mathbf{X}^{\scriptscriptstyle{+}}_{\vec{\mu}}(P_{m,-n})\,V_{\alpha}\,\mathbf{X}_{\vec{\lambda}}(P)|P\rangle}
   {\langle P_{m,-n}|V_{\alpha}|P\rangle}=\frac{\langle P_{m,-n}|(X^{(m,n)}_{\nu})^{\scriptscriptstyle{+}}\,V_{\alpha}\,\mathbf{X}_{\vec{\lambda}}(P)|P\rangle}
   {\langle P_{m,-n}|V_{\alpha}|P\rangle}=\\=
   \frac{\langle P_{m,n}|D_{m,n}^{+}\,(X^{(m,n)}_{\nu})^{\scriptscriptstyle{+}}\,V_{\alpha}\,\mathbf{X}_{\vec{\lambda}}(P)|P\rangle}
   {\langle P_{m,n}|V_{\alpha}|P\rangle}
   \frac{\langle P_{m,n}|V_{\alpha}|P\rangle}{\langle P_{m,n}|D_{m,n}^{+}V_{\alpha}|P\rangle}.
\end{multline}
Using \eqref{Pmn-polynomial}, conjugated version of \eqref{XX-def} and \eqref{Jack-General-matrix-element} we can rewrite the r.h.s. of \eqref{XX-prop-proof-1} as
\begin{multline}
  \frac{\langle P_{m,n}|D_{m,n}^{+}\,(X^{(m,n)}_{\nu})^{\scriptscriptstyle{+}}\,V_{\alpha}\,\mathbf{X}_{\vec{\lambda}}(P)|P\rangle}
   {\langle P_{m,n}|V_{\alpha}|P\rangle}
   \frac{\langle P_{m,n}|V_{\alpha}|P\rangle}{\langle P_{m,n}|D_{m,n}^{+}V_{\alpha}|P\rangle}=\\=
   (-1)^{mn}\mathbb{P}^{-1}_{m,n}(\alpha,P)
   \frac{_{\nu,\scriptscriptstyle{\varnothing}}\langle P_{m,n}|V_{\alpha}\,\mathbf{X}_{\vec{\lambda}}(P)|P\rangle}
   {\langle P_{m,n}|V_{\alpha}|P\rangle}\overset{\eqref{Jack-General-matrix-element}}{=}
   (-1)^{mn}\mathbb{P}^{-1}_{m,n}(\alpha,P)
    Z_{\text{\sf bif}}(\alpha|P_{m,n},(\nu,\varnothing);P,\vec{\lambda}).
\end{multline}
And finally using the identity (see Appendix \ref{ME-reduction})
\begin{equation}\label{Z-Z-relat-2}
    Z_{\text{\sf bif}}(\alpha|P_{m,n},(\nu,\varnothing);P,\vec{\lambda})=(-1)^{mn}\,\mathbb{P}_{m,n}(\alpha,P)\,
    Z_{\text{\sf bif}}(\alpha|P_{m,-n},(\mu_{1},\mu_{2});P,\vec{\lambda}),
\end{equation}
we arrive at \eqref{General-General-matrix-element} for $P'=P_{m,-n}$ for some infinite set of integers $m$ and $n$. Since the matrix element in \eqref{General-General-matrix-element} is a rational function in $P'$ the proof of Proposition \ref{XX-prop} is complete. $\square$
\begin{corollary}\label{norm-Corollary}
The states $\mathbf{X}_{\vec{\lambda}}(P)|P\rangle$ form an orthogonal basis
\begin{equation}
    \langle P|\mathbf{X}^{\scriptscriptstyle{+}}_{\vec{\mu}}(P)\mathbf{X}_{\vec{\lambda}}(P)|P\rangle=
    \mathcal{N}_{\vec{\lambda}}(P)\times\delta_{\vec{\lambda},\vec{\mu}},
\end{equation}
where $\delta_{\vec{\lambda},\vec{\mu}}=0$ if $\vec{\lambda}\neq\vec{\mu}$, $\delta_{\vec{\lambda},\vec{\lambda}}=1$ and 
\begin{equation}\label{norm}
   \mathcal{N}_{\vec{\lambda}}(P)=1/Z_{\text{\sf{vec}}}(P,\vec{\lambda}).
\end{equation}
\end{corollary}
\paragraph{Proof of Corollary \ref{norm-Corollary}:}First, we note that the states $\mathbf{X}_{\vec{\lambda}}(P)|P\rangle$ and $\langle P|\mathbf{X}^{\scriptscriptstyle{+}}_{\vec{\mu}}(P)$ are trivially orthogonal for different levels i.e. for $|\vec{\lambda}|\neq|\vec{\mu}|$. So, we have to check the orthogonality for the states from the same level. It is easy to see that
\begin{equation}\label{scalar-product-matrix-element}
  \langle P|\mathbf{X}^{\scriptscriptstyle{+}}_{\vec{\mu}}(P)\mathbf{X}_{\vec{\lambda}}(P)|P\rangle=
   Z_{\text{\sf bif}}(0|P,\vec{\mu};P,\vec{\lambda})\quad
   \text{for}\quad |\vec{\lambda}|=|\vec{\mu}|.
\end{equation}
Analyzing the explicit form \eqref{Zbif-def} of the function $Z_{\text{\sf bif}}(0|P,\vec{\mu};P,\vec{\lambda})$ one can show that it vanishes always except for $\vec{\lambda}=\vec{\mu}$. Using \eqref{Zvec-def} we arrive at  \eqref{norm}. $\square$

Thus we proved the existence of the orthogonal basis which satisfies \eqref{matrix-elements}. In order to complete the proof of Proposition \ref{MAIN-Proposition} we have to prove its uniqueness.
\begin{Proposition}\label{Unique-Prop}
The basis of states $\mathbf{X}_{\vec{\lambda}}(P)|P\rangle$ defined above is the unique basis satisfying
\begin{equation}\label{General-General-matrix-element-final}
     \frac{\langle P'|\mathbf{X}^{\scriptscriptstyle{+}}_{\vec{\mu}}(P')\,V_{\alpha}\,\mathbf{X}_{\vec{\lambda}}(P)|P\rangle}{\langle P'|V_{\alpha}|P\rangle}=
     Z_{\text{\sf bif}}(\alpha|P',\vec{\mu};P,\vec{\lambda}).
\end{equation}
\end{Proposition}
\paragraph{Proof of Proposition \ref{Unique-Prop}:}The proof is based on induction. Assume that for given $N$ the set of  states $\mathbf{X}_{\vec{\lambda}}(P)|P\rangle$ with $|\vec{\lambda}|\leq N$ is unique. Then suppose that at the level $N+1$ there are two states $\mathbf{X}_{\vec{\lambda}}(P)|P\rangle$ and $\tilde{\mathbf{X}}_{\vec{\lambda}}(P)|P\rangle$ for some partition $\vec{\lambda}$ which have the same matrix elements with all the states $\langle P'|\mathbf{X}^{\scriptscriptstyle{+}}_{\vec{\mu}}(P')$ with $|\vec{\mu}|\leq N$, i.e.
\begin{equation}\label{X-Xtilde-matrix-element}
    \frac{\langle P'|\mathbf{X}^{\scriptscriptstyle{+}}_{\vec{\mu}}(P')\,V_{\alpha}\,\mathbf{X}_{\vec{\lambda}}(P)|P\rangle}{\langle P'|V_{\alpha}|P   \rangle}=
    \frac{\langle P'|\mathbf{X}^{\scriptscriptstyle{+}}_{\vec{\mu}}(P')\,V_{\alpha}\,\tilde{\mathbf{X}}_{\vec{\lambda}}(P)|P\rangle}{\langle P'|V_{\alpha}|P\rangle}.
\end{equation}
Also define the operator 
\begin{equation*}
  Y_{N+1}(P)=\mathbf{X}_{\vec{\lambda}}(P)-\tilde{\mathbf{X}}_{\vec{\lambda}}(P).
\end{equation*}
Then as follows from \eqref{X-Xtilde-matrix-element}
\begin{equation}\label{X-null-condition}
    \langle P'|\mathbf{X}^{\scriptscriptstyle{+}}_{\vec{\mu}}(P')\,V_{\alpha}\,Y_{N+1}(P)|P\rangle=0\qquad\text{for any}\;\vec{\mu}:
    \;\;|\vec{\mu}|\leq N.
\end{equation}
Since the states $\langle P'|\mathbf{X}^{\scriptscriptstyle{+}}_{\vec{\mu}}(P')$ form a basis \eqref{X-null-condition} is equivalent to
\begin{equation}\label{X-null-condition-2}
    \langle P'|\hat{L}_{\mu_{1}}\hat{a}_{\mu_{2}}\,V_{\alpha}\,Y_{N+1}(P)|P\rangle=0\qquad\text{for any}\;\mu_{1}, \mu_{2}:
    \;\;|\mu_{1}|+|\mu_{2}|\leq N,
\end{equation}
which by Lemma \ref{lemma} implies $Y_{N+1}=0$. $\square$

The proof of Proposition \ref{MAIN-Proposition} is now complete. We have shown that the basis $|P\rangle_{\vec{\lambda}}\overset{\text{def}}{=}\mathbf{X}_{\vec{\lambda}}(P)|P\rangle$ exists and unique. We would like to comment how our construction can be used to actually define the basis states $|P\rangle_{\vec{\lambda}}$. As was claimed in Proposition \ref{Jack-Proposition} the states with second empty diagram  $|P\rangle_{\lambda,\scriptscriptstyle{\varnothing}}$ are given by \eqref{Jack-basis}. The rest of the basis can be constructed by \eqref{XX-def} and \eqref{Jack-General-matrix-element-2}. Since the coefficients $C_{\vec{\lambda}}^{\vec{\mu}}(P)$ in \eqref{X-General} are rational functions of the momenta $P$ they can be restored completely from the set of equations \eqref{XX-def}. Indeed, using  \eqref{Jack-General-matrix-element-2} equation  \eqref{XX-def} can be written as
\begin{equation}\label{XX-def-2}
   |P_{m,n}\rangle_{\lambda,\scriptscriptstyle{\varnothing}}=(-1)^{mn}\mathbf{X}_{(\lambda_{1},\lambda_{2})}(P_{m,-n})\,D_{m,n}|P_{m,n}\rangle,
\end{equation}
where the pair $(\lambda_{1},\lambda_{2})$ is defined from the partition $\lambda$ by the following rule
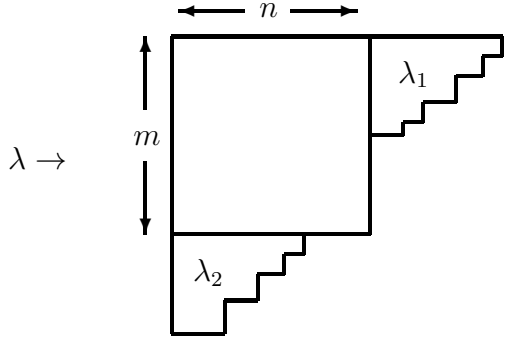
\begin{equation*}
    \begin{picture}(50,140)(60,-120)
    \Thicklines
    \unitlength 2.5pt 
    \rotatebox{270}{
    \put(0,50){\line(0,1){20}}
    \put(0,50){\line(1,0){15}}
    \put(15,50){\line(0,1){5}}
    \put(0,70){\line(1,0){3}}
    \put(3,70){\line(0,-1){3}}
    \put(3,67){\line(1,0){3}}
    \put(6,67){\line(0,-1){4}}
    \put(6,63){\line(1,0){4}}
    \put(10,63){\line(0,-1){5}}
    \put(10,58){\line(1,0){3}}
    \put(13,58){\line(0,-1){3}}
    \put(13,55){\line(1,0){2}}
    \put(30,20){\line(0,1){20}}
    \put(30,20){\line(1,0){15}}
    \put(45,20){\line(0,1){5}}
    \put(30,40){\line(1,0){3}}
    \put(33,40){\line(0,-1){3}}
    \put(33,37){\line(1,0){3}}
    \put(36,37){\line(0,-1){4}}
    \put(36,33){\line(1,0){4}}
    \put(40,33){\line(0,-1){5}}
    \put(40,28){\line(1,0){5}}
    \put(45,28){\line(0,-1){3}}
    \put(0,20){\line(0,1){40}}
    \put(0,20){\line(1,0){40}}
    \put(30,20){\line(0,1){30}}
    \put(30,50){\line(-1,0){30}}
    \put(19,16){\vector(1,0){11}}
    \put(11,16){\vector(-1,0){11}}
    \put(-3.8,38){\vector(0,1){10}}
    \put(-3.8,31){\vector(0,-1){10}}
    }
    \put(-75,-20){\mbox{$\lambda\rightarrow$}}
    \put(-16,-7){\mbox{$\lambda_{1}$}}
    \put(-47,-37){\mbox{$\lambda_{2}$}}
    \put(-56,-16.5){\mbox{$m$}}
    \put(-37,3){\mbox{$n$}}
    \end{picture}
\end{equation*}
We note that for given pair of Young diagrams $(\lambda_{1},\lambda_{2})$ the equation \eqref{XX-def-2} is valid for any ``master'' diagram $\lambda$ which ``fits'' $(\lambda_{1},\lambda_{2})$ as shown on the picture.  It allows to compute the coefficients $C_{\vec{\lambda}}^{\vec{\mu}}(P)$ in $\mathbf{X}_{\vec{\lambda}}(P)$ for $P=P_{m,-n}$ for some infinite set of integers $m$ and $n$ and hence to restore them completely.
In fact, it can be proven that all of them are polynomials in $P$:
\begin{corollary}
All the coefficients $C_{\vec{\lambda}}^{\vec{\mu}}(P)$ in 
\begin{equation}
   \mathbf{X}_{\vec{\lambda}}(P)=\sum_{|\vec{\mu}|=|\vec{\lambda}|}C_{\vec{\lambda}}^{\vec{\mu}}(P)\hat{a}_{-\mu_{1}}\hat{L}_{-\mu_{2}},
\end{equation}
are polynomials of the momenta $P$.
\end{corollary}
\section{Concluding remarks}\label{Remarks}
We note that the ``$U(1)$ factor'' in the r.h.s. in \eqref{conformal-block-refined} can be interpreted up to some trivial factors as a  free field correlation function of chiral vertex operators \eqref{vertex}. Namely, one has
\begin{equation}\label{free-field-correlation-function}
   \langle\mathcal{V}_{\alpha_{1}}(z_{1})\dots\mathcal{V}_{\alpha_{n}}(z_{n})\rangle=
   \prod_{i<j}\left(1-\frac{z_{j}}{z_{i}}\right)^{2\alpha_{i}(Q-\alpha_{j})},
\end{equation}
and hence naively the ``dressing'' of Virasoro primaries $V_{\alpha}^{L}$ by the free-field vertex operators $\mathcal{V}_{\alpha}$ multiplies the conformal block by the  trivial factor \eqref{free-field-correlation-function}. But in fact the r\^ole of auxiliary bosonic field justifies itself when one expands the conformal block over the intermediate  states. We note that the states $|P\rangle_{\vec{\lambda}}$ have contributions from different levels of Verma module for Virasoro algebra and the ``mixing'' of levels is governed by this auxiliary bosonic field.

The generalization of  our construction to other algebras will be discussed  in a future publication.  We stress here that there is an analog  of the infinite system of commuting quantities considered in Appendix \ref{IM} for $W_{n}$ algebras. The matrix elements of certain primary fields between its eigenstates also have a nice factorized form. This should explain the AGT conjecture for higher rank groups \cite{Wyllard:2009hg,Mironov:2009by}. For example for $W_{3}$ algebra (the algebra generated by two holomorphic currents $T(z)$ of spin $2$ and $W(z)$ of spin $3$) this system starts with the first non-trivial integral
\begin{equation}
\mathbf{I}=\sqrt{\frac{8}{3}}\left(6iQ\sqrt{\frac{3}{8}}\sum_{k>0}ka_{-k}a_{k} +\sum_{k\neq0}a_{-k}L_{k}+
\frac{\sqrt{4+15Q^{2}}}{4}W_{0}+\frac{1}{3}\sum_{i+j+k=0}a_{i}a_{j}a_{k}\right),  \label{I}%
\end{equation}
where $a_{k}$ are again generators of an auxiliary Heisenberg algebra. The eigenstates of the integral \eqref{I} have very simple form and their eigenvalues are linear functions of the momenta. The matrix elements of certain primary field between these eigenstates have again factorized form coinciding  with $Z_{\text{\sf bif}}$ for $U(3)$ linear quiver gauge theories\footnote{We have checked it up to level two.}. We will make this statement more precise elsewhere. We comment here that the generalization of our construction for $WA_{n}$ algebras will require a version of $A_{n}$ Selberg integral \cite{Warnaar-An} with insertion of two Jack polynomials.
\section*{Acknowledgments}
The idea about the existence of an orthogonal basis diagonalizing an infinite system of Integrals of Motion in algebra $\mathcal{A}$ which can explain the AGT conjecture was proposed by Boris Feigin in a more general context. We thank him for sharing his insights with us. 

The authors are grateful to Alexander Belavin, Mikhail Bershtein, Sergei Parkhomenko, Yaroslav Pugai and Jun'ichi Shiraishi for numerous discussions. A.~L. thanks Sergei Lukyanov and Alexander Zamolodchikov for support and interest to this work. He also thanks the organizers of the workshop ``From Sigma Models to Four-dimensional QFT'' DESY, December 2010 and especially J\"org Teschner for hospitality during the last stage of this project.

This research was held within the framework of the Federal programs ``Scientific and Scientific-Pedagogical Personnel of Innovational Russia'' on 2009-2013 (state contracts No. P1339 and No. 02.740.11.5165) and was supported by cooperative CNRS-RFBR  grant PICS-09-02-93064,  by RFBR  initiative interdisciplinary project grant 11-01-12023-OFI-m  and by Russian Ministry of Science and Technology under the Scientific Schools grant 6501.2010.2. The research of V.~A. is also supported by RFBR grant 10-02-00499 and Goskontrakt No. 14.740.11.0081.

\paragraph{Notes added in proof:}We thank Referees for their valuable  comments and suggestions which lead to the substantial improvements in the exposition of the paper.
\Appendix
\section{Selberg integral with insertion of two Jack polynomials}\label{Selberg-proof}
In this appendix we compute the integral appearing  in the r.h.s. in \eqref{Generalized-Selberg}. It is equivalent to the computation of the following integral which generalizes both Kadell \cite{kadell} and Hua-Kadell \cite{kadell-2,Hua,Forrester:2007fk} integrals. Namely, let $\lambda$ and $\mu$ partitions such that $\lambda$ is of length at most $n$. Then define 
\begin{equation}\label{Gen-Selberg-integral}
    I_{\lambda,\mu}^{(n)}(\alpha,\beta,\gamma)=\frac{1}{n!}\int\limits_{[0,1]^{n}}
    \mathbf{P}_{\lambda}^{\scriptscriptstyle{(1/\gamma)}}[p_{k}]\,
    \mathbf{P}_{\mu}^{\scriptscriptstyle{(1/\gamma)}}[p_{k}+(\beta-\gamma)/\gamma]\,
    \,\prod_{j=1}^{n}t_{j}^{\alpha-1}(1-t_{j})^{\beta-1}\prod_{i<j}|t_{i}-t_{j}|^{2\gamma}
    dt_{1}\dots dt_{n},
\end{equation}
where integration goes over $n-$dimensional cube $t_{i}\in[0,1]$ for $i=1,\dots,n$ and the  parameters $\alpha$, $\beta$ and $\gamma$ are subject to  conditions which ensure convergence of the integral \eqref{Gen-Selberg-integral}. Polynomial
$\mathbf{P}_{\lambda}^{\scriptscriptstyle{(1/\gamma)}}[p_{k}]=\mathbf{P}_{\lambda}^{\scriptscriptstyle{(1/\gamma)}}\left(t_{1},\dots,t_{n}\right)$ is the Jack polynomial normalized as \cite{Macdonald}
\begin{equation*}
   \mathbf{P}_{\lambda}^{\scriptscriptstyle{(1/\gamma)}}\left(t_{1},\dots,t_{n}\right)=m_{\lambda}(t)+\sum_{\mu\prec\mu}u_{\lambda,\mu}\,m_{\mu}(t),
\end{equation*}
where $m_{\lambda}(t)$ is the monomial symmetric polynomial and the sum goes over partitions $\mu$ dominated by $\lambda$.
We note that in section \ref{Proof-section} we used different normalization of Jack polynomials. They are related as
\begin{equation}\label{Jack-different-norm}
    \jac_{\lambda}^{\scriptscriptstyle{(1/\gamma)}}(t_{1},\dots,t_{n})=
    \mathsf{c}_{\lambda}(\gamma)\,
    \mathbf{P}_{\lambda}^{\scriptscriptstyle{(1/\gamma)}}\left(t_{1},\dots,t_{n}\right),
\end{equation}
with 
\begin{equation*}
   \mathsf{c}_{\lambda}(\gamma)=\prod_{s\in\lambda}\left(1+l_{\lambda}(s)+\gamma^{-1} a_{\lambda}(s)\right)
\end{equation*}
where $a_{\lambda}(s)$ and $l_{\mu}(s)$ are correspondingly the arm length and the leg length of the square $s$ in the partition $\lambda$. We propose that the integral \eqref{Gen-Selberg-integral} can be computed exactly\footnote{We note that for $\mu=\varnothing$ it coincides with Kadell integral \cite{kadell} while for $\beta=\gamma$ and both general diagrams $\lambda$ and $\mu$ with Hua-Kadell integral \cite{kadell-2,Hua}.}. Before proceed further we comment how the integral  \eqref{Gen-Selberg-integral} can be related to the integral in the r.h.s. in \eqref{Generalized-Selberg}. We use the following identity proved by Kadell \cite{kadell}
\begin{equation}\label{Kadell-identity}
     \mathbf{P}_{\lambda}^{\scriptscriptstyle{(1/\gamma)}}[p_{-k}]=\prod_{j=1}^{n}t_{j}^{-\lambda_{1}}\,
     \mathbf{P}_{\hat{\lambda}}^{\scriptscriptstyle{(1/\gamma)}}[p_{k}],
\end{equation}
where for given diagram $\lambda=\{\lambda_{1}\geq \lambda_{2}\geq\dots\}$ the ``hatted'' diagram $\hat{\lambda}=\{\hat{\lambda}_{1}\geq \hat{\lambda}_{2}\geq\dots\}$ is defined by
\begin{equation}\label{lambda-hat}
     \hat{\lambda}_{j}=\lambda_{1}-\lambda_{n-j+1}.
\end{equation}
Then the integral in the r.h.s. \eqref{Generalized-Selberg} can be related to the integral \eqref{Gen-Selberg-integral} as
\begin{equation}\label{Sel-Sel-relation}
    \left\langle\jac_{\mu}^{\scriptscriptstyle{(1/g)}}[p_{k}+\rho]\,\jac_{\lambda}^{\scriptscriptstyle{(1/g)}}[p_{-k}]\right\rangle_{\text{\sf Sel}}^{(n)}=
    \mathsf{c}_{\lambda}(g)\mathsf{c}_{\mu}(g)\,
    I_{\hat{\lambda},\mu}^{(n)}(1+A-\lambda_{1},1+B,g),
\end{equation}
where $\rho=(1+B-g)/g$.
For the future convenience we also introduce the generalized Pochhammer symbol
\begin{equation}\label{Gen-Pochhammer}
    [x]_{\lambda}=\prod_{j\geq1}(x+(1-j)\gamma)_{\lambda_{j}}=\prod_{j\geq 1}\frac{\Gamma(x+\lambda_{j}+(1-j)\gamma)}{\Gamma(x+(1-j)\gamma)}.
\end{equation}

Our method of computation of the integral \eqref{Gen-Selberg-integral} is a step-by-step copy of the evaluation of the Kadell integral given in \cite{warnaar-2008-40}\footnote{We note that the method suggested in \cite{warnaar-2008-40} is a generalization of Anderson's derivation of Selberg integral \cite{Anderson} (see also \cite{Forrester}). Similar method was applied to ``complex'' version of Selberg like integrals in \cite{Fateev:2007qn,Fateev:2007ab,Fateev:2008bm}}  and will be based on two integral  identities. The first one is attributed to Okounkov and Olshanski \cite{Okounkov:fk} (see also \cite{Kuznetsov:2003fk}):
\begin{identity}\label{OO-1}
Let $\tau=(\tau_{1},\dots,\tau_{n-1})$ and $t=(t_{1},\dots,t_{n})$ satisfy the interlacing property 
\begin{equation*}
    t_{1}<\tau_{1}<t_{2}<\tau_{2}<\dots<t_{n-1}<\tau_{n-1}<t_{n}
\end{equation*}
denoted by $\tau\prec t$. Then for $\nu=(\nu_{1}\geq \nu_{2}\geq\dots)$ a partition of length at most $n-1$
\begin{multline}\label{OO-1-Int}
    \prod_{i<j=1}^{n}(t_{j}-t_{i})^{2\gamma-1}\,\mathbf{P}_{\nu}^{\scriptscriptstyle{(1/\gamma)}}\left(t_{1},\dots,t_{n}\right)=
    \Lambda_{\nu}(\gamma)\times\\\times
    \int\limits_{\tau\prec t}\mathbf{P}_{\nu}^{\scriptscriptstyle{(1/\gamma)}}\left(\tau_{1},\dots,\tau_{n-1}\right)\,
    \prod_{i<j=1}^{n-1}(\tau_{j}-\tau_{i})\,
    \prod_{i=1}^{n-1}\prod_{j=1}^{n}|\tau_{i}-t_{j}|^{\gamma-1}\,d\tau_{1}\dots d\tau_{n-1},
\end{multline}
where 
\begin{equation*}
    \Lambda_{\nu}(\gamma)=\frac{\Gamma(n\gamma)}{\Gamma^{n}(\gamma)}\,
    \frac{[n\gamma]_{\nu}}{[(n-1)\gamma]_{\nu}},
\end{equation*}
and $\Gamma(x)$ is Euler Gamma-function.
\end{identity} 
The second identity states:
\begin{identity}\label{OO-2}
Let  $t=(t_{1},\dots,t_{n})$ and $\tau=(0,\tau_{1},\dots,\tau_{n-1},1)$ satisfy the interlacing property
\begin{equation*}
    0<t_{1}<\tau_{1}<t_{2}<\tau_{2}<\dots<t_{n-1}<\tau_{n-1}<t_{n}<1
\end{equation*}
denoted by $t\prec\tau$. Then for $\mu=(\mu_{1}\geq \mu_{2}\geq\dots)$ a partition
\begin{multline}\label{OO-2-Int}
   \int\limits_{t\prec\tau}\mathbf{P}_{\mu}^{\scriptscriptstyle{(1/\gamma)}}[p_{k}(t)+(\beta-\gamma)/\gamma]\,
   \prod_{i<j=1}^{n}(t_{j}-t_{i})\,\prod_{j=1}^{n}t_{j}^{\alpha-1}(1-t_{j})^{\beta-1}\,
    \prod_{i=1}^{n}\prod_{j=1}^{n-1}|t_{i}-\tau_{j}|^{\gamma-1}\,dt_{1}\dots dt_{n}=\\=
    \Xi_{\mu}(\alpha,\beta,\gamma)\,
    \prod_{j=1}^{n-1}\tau_{j}^{\alpha+\gamma-1}(1-\tau_{j})^{\beta+\gamma-1}
    \prod_{i<j=1}^{n-1}(\tau_{j}-\tau_{i})^{2\gamma-1}\,
    \mathbf{P}_{\mu}^{\scriptscriptstyle{(1/\gamma)}}[p_{k}(\tau)+\beta/\gamma],
\end{multline}
where 
\begin{equation*}
    \Xi_{\mu}(\alpha,\beta,\gamma)=\frac{\Gamma(\alpha)\Gamma(\beta)\Gamma^{n-1}(\gamma)}{\Gamma(\alpha+\beta+(n-1)\gamma)}\,
    \frac{[\alpha+\beta+(n-2)\gamma]_{\mu}}{[\alpha+\beta+(n-1)\gamma]_{\mu}}.
\end{equation*}
\end{identity}
We note that the Identity \ref{OO-2} follows immediately from the Identity \ref{OO-1} for $\alpha=N\gamma$ and $\beta=M\gamma$ for non-negative integers $M$ and $N$. Indeed, let us take the number of integrations $(n-1)$ in \eqref{OO-1-Int} larger enough than the length of the diagram $\lambda$. Then we can consider the common limit    $t_{1},\dots,t_{N}\rightarrow0$ and $t_{n-M+1},\dots,t_{n}\rightarrow1$ in \eqref{OO-1-Int}\footnote{The taking of this limit is not quite hard task, but few comments might be useful. Taking $t_{1}\rightarrow0$ in \eqref{OO-1-Int} is not singular and hence we just put $t_{1}=0$ in \eqref{OO-1-Int}. Then taking $t_{2}\rightarrow0$ we see that the integration variable $\tau_{1}$ is confined between two points $0$ and $t_{2}$ approaching each other. Such an integral is singular and the leading singularity can be easily computed $\int_{0}^{t_{2}}\tau_{1}^{\gamma-1}(\tau_{1}-t_{2})^{\gamma-1}d\tau_{1}=t_{2}^{2\gamma-1}\Gamma^{2}(\gamma)/\Gamma(2\gamma)$ which has the same behavior at $t_{2}\rightarrow0$ as  the l.h.s. of \eqref{OO-1-Int}. Taking $t_{3}\rightarrow0$ and further as well as $t_{n-m+i}\rightarrow1$, $i=1,\dots,M$ are similar.}. Using the fact that
\begin{equation*}
    \mathbf{P}_{\lambda}^{\scriptscriptstyle{(1/\gamma)}}\bigl(\underbrace{0,\dots,0}_{N},t_{1},\dots,t_{n}\bigr)=
    \mathbf{P}_{\lambda}^{\scriptscriptstyle{(1/\gamma)}}\bigl(t_{1},\dots,t_{n}\bigr)\quad\text{and}\quad
     \mathbf{P}_{\lambda}^{\scriptscriptstyle{(1/\gamma)}}\bigl(t_{1},\dots,t_{n},\underbrace{1,\dots,1}_{M}\bigr)=
    \mathbf{P}_{\lambda}^{\scriptscriptstyle{(1/\gamma)}}[p_{k}(t)+M],
\end{equation*}
we arrive at \eqref{OO-2-Int} with $\alpha=N\gamma$ and $\beta=M\gamma$, $M,N\in\mathbb{Z}_{\geq0}$. So, we propose \eqref{OO-2-Int} as a continuation from integer values of $M$ and $N$. Such an ``analytic'' continuation typically  requires the use of Carlson's theorem \cite{Carlson}. The more rigorous proof of \eqref{OO-2-Int} can be probably done within  the lines of \cite{Kuznetsov:2003fk}.

We now proceed to compute the integral \eqref{Gen-Selberg-integral}. Before do that, let us use the symmetry of the integrand in \eqref{Gen-Selberg-integral} and change
\begin{equation*}
    \frac{1}{n!}\,\int\limits_{[0,1]^{n}}\longrightarrow\;\idotsint\limits_{0<t_{1}<t_{2}<\dots<t_{n}<1}
\end{equation*}
We can reduce the number of integrations in \eqref{Gen-Selberg-integral} performing the following steps:
\begin{itemize}
\item First, for $\lambda=(\lambda_{1},\dots,\lambda_{n})$ we use the formula
   \begin{equation*}
         \mathbf{P}_{\lambda}^{\scriptscriptstyle{(1/\gamma)}}[p_{k}(t)]=(t_{1}\dots t_{n})^{\lambda_{n}}\,
         \mathbf{P}_{\nu}^{\scriptscriptstyle{(1/\gamma)}}[p_{k}(t)],
   \end{equation*}
   where $\nu=(\lambda_{1}-\lambda_{n},\dots,\lambda_{n-1}-\lambda_{n},0)$
\item Second, we represent
   \begin{equation*}
       \prod_{i<j=1}^{n}(t_{j}-t_{i})^{2\gamma-1}\,\mathbf{P}_{\nu}^{\scriptscriptstyle{(1/g)}}\left(t_{1},\dots,t_{n}\right)
   \end{equation*}
in \eqref{Gen-Selberg-integral} using \eqref{OO-1-Int}. 
\item And finally, we compute the remaining integral
   \begin{equation*}
      \int\limits_{t\prec\tau}\mathbf{P}_{\mu}^{\scriptscriptstyle{(1/\gamma)}}[p_{k}(t)+(\beta-\gamma)/\gamma]\,
      \prod_{i<j=1}^{n}(t_{j}-t_{i})\,\prod_{j=1}^{n}t_{j}^{\alpha+\lambda_{n}-1}(1-t_{j})^{\beta-1}\,
    \prod_{i=1}^{n}\prod_{j=1}^{n-1}|t_{i}-\tau_{j}|^{\gamma-1}\,dt_{1}\dots dt_{n}
   \end{equation*}
   using \eqref{OO-2-Int}.
\end{itemize}
As a result we reduced our original integral \eqref{Gen-Selberg-integral}  to the integral of the same form, but with lower number of integrations $n\rightarrow n-1$
\begin{multline}\label{Gen-Selberg-integral-recursion} 
   I_{\lambda,\mu}^{(n)}(\alpha,\beta,\gamma)=\frac{[n\gamma]_{\nu}}{[(n-1)\gamma]_{\nu}}
   \frac{\Gamma(\alpha+\lambda_{n})\Gamma(\beta)\Gamma(n\gamma)}{\Gamma(\alpha+\beta+(n-1)\gamma+\lambda_{n})\Gamma(\gamma)}
   \frac{[\alpha+\beta+(n-2)\gamma+\lambda_{n}]_{\mu}}{[\alpha+\beta+(n-1)\gamma+\lambda_{n}]_{\mu}}\times\\\times
   I_{\nu,\mu}^{(n-1)}(\alpha+\gamma,\beta+\gamma,\gamma),
\end{multline}
where $\nu=(\lambda_{1}-\lambda_{n},\dots,\lambda_{n-1}-\lambda_{n},0)$. Since the right-hand side in \eqref{Gen-Selberg-integral-recursion} is again the integral of the form  \eqref{Gen-Selberg-integral} we can proceed further by induction. Using
\begin{equation*}
     \frac{[n\gamma]_{\nu}}{[(n-1)\gamma]_{\nu}}=\frac{\mathbf{P}_{\lambda}^{\scriptscriptstyle{1/\gamma}}[n]}
     {\mathbf{P}_{\nu}^{\scriptscriptstyle{1/\gamma}}[n-1]},
\end{equation*}
we find
\begin{multline}\label{Sel-answer}
   I_{\lambda,\mu}^{(n)}(\alpha,\beta,\gamma)=\mathbf{P}_{\lambda}^{\scriptscriptstyle{1/\gamma}}[n]\,
   \mathbf{P}_{\mu}^{\scriptscriptstyle{1/\gamma}}[n+(\beta-\gamma)/\gamma]\,
   \prod_{j=1}^{n}\frac{\Gamma(\alpha+(n-j)\gamma+\lambda_{j})\Gamma(\beta+(j-1)\gamma)\Gamma(j\beta)}
   {\Gamma(\alpha+\beta+(2n-j-1)\gamma+\lambda_{j})\Gamma(\gamma)}\times\\\times
   \prod_{j=1}^{n}\frac{[\alpha+\beta+(2n-j-2)\gamma+\lambda_{j}]_{\mu}}{[\alpha+\beta+(2n-j-1)\gamma+\lambda_{j}]_{\mu}}.
\end{multline}
Now, as was claimed in \eqref{Sel-Sel-relation}
\begin{equation}
    \frac{\left\langle\jac_{\mu}^{\scriptscriptstyle{(1/g)}}[p_{k}+\rho]\,\jac_{\lambda}^{\scriptscriptstyle{(1/g)}}[p_{-k}]\right\rangle_{\text{\sf Sel}}^{(n)}}
    {\left\langle1\right\rangle_{\text{\sf Sel}}^{(n)}}=
     \mathsf{c}_{\lambda}(g)\mathsf{c}_{\mu}(g)\,
    \frac{I_{\hat{\lambda},\mu}^{(n)}(1+A-\lambda_{1},1+B,g)}{I_{\scriptscriptstyle{\varnothing,\varnothing}}^{(n)}(1+A,1+B,g)},
\end{equation}
where the partition $\hat{\lambda}$ is defined by \eqref{lambda-hat}. Using evaluation formula \cite{Stanley}
\begin{equation}
    \mathbf{P}_{\lambda}^{\scriptscriptstyle{(1/\gamma)}}[N]=\gamma^{-|\lambda|}\frac{[n\gamma]_{\lambda}}{\mathsf{c}_{\lambda}(\gamma)},
\end{equation}
and \eqref{Sel-answer} we find
\begin{multline}\label{Sel-answer-2}
    \frac{\left\langle\jac_{\mu}^{\scriptscriptstyle{(1/g)}}[p_{k}+\rho]\,\jac_{\lambda}^{\scriptscriptstyle{(1/g)}}[p_{-k}]\right\rangle_{\text{\sf Sel}}^{(n)}}
    {\left\langle1\right\rangle_{\text{\sf Sel}}^{(n)}}=
    g^{-|\lambda|-|\mu|}[ng]_{\lambda}[1+B+(n-1)g]_{\mu}\times\\\times
   \prod_{j=1}^{n}\frac{\Gamma(1+A+(j-1)g-\lambda_{j})}{\Gamma(1+A+(j-1)g)}
   \frac{\Gamma(2+A+B+(n+j-2)g)}{\Gamma(2+A+B+(n+j-2)g-\lambda_{j})}
   \frac{[2+A+B+(n+j-3)g-\lambda_{j}]_{\mu}}{[2+A+B+(n+j-2)g-\lambda_{j}]_{\mu}}
\end{multline}
Note that if $m$ is any integer exceeding the length of $\mu$ (i.e. $l(\mu)\leq m$) we can also rewrite the r.h.s. of  \eqref{Sel-answer-2} as
\begin{multline}\label{Sel-answer-3}
    g^{-|\lambda|-|\mu|}[ng]_{\lambda}[1+B+(n-1)g]_{\mu}\prod_{j=1}^{n}\frac{\Gamma(1+A+(j-1)g-\lambda_{j})}{\Gamma(1+A+(j-1)g)}
    \prod_{j=1}^{m}\frac{\Gamma(2+A+B+(2n-1-j)g)}{\Gamma(2+A+B+(2n-1-j)g+\mu_{j})}
    \times\\\times
    \frac{\prod_{i=1}^{n+1}\prod_{j=1}^{m}\Gamma(2+A+B+\mu_{j}-\lambda_{i}+(n+i-j-2)g)}
    {\prod_{i=1}^{n}\prod_{j=1}^{m+1}\Gamma(2+A+B+\mu_{j}-\lambda_{i}+(n+i-j-1)g)}
    \frac{\prod_{i=1}^{n}\Gamma(2+A+B+(n+j-2)g)}{\prod_{j=1}^{m}\Gamma(2+A+B+(2n-j-1)g)}
\end{multline}
Using \eqref{Omega}, \eqref{ABg}, \eqref{Zbif-dual} and \eqref{Zbif-dual-2} we find 
\begin{equation}
    \frac{\left\langle\jac_{\mu}^{\scriptscriptstyle{(1/g)}}[p_{k}+\rho]\,\jac_{\lambda}^{\scriptscriptstyle{(1/g)}}[p_{-k}]\right\rangle_{\text{\sf Sel}}^{(n)}}
    {\langle1\rangle_{\text{\sf Sel}}^{(n)}}=\Omega^{-1}_{\lambda}(P)\Omega^{-1}_{\mu}(P')
    Z_{\text{\sf{bif}}}(\alpha_{n}|P',(\mu,\varnothing);P,(\lambda,\varnothing)),
\end{equation}
which implies \eqref{Selberg-matrix-element}.
\section{Proof of the identities \eqref{Z-Z-relat} and \eqref{Z-Z-relat-2}}\label{ME-reduction}
We note that the function $Z_{\text{\sf{bif}}}(\alpha|P',\vec{\mu};P,\vec{\lambda})$ defined by \eqref{Zbif-def} can be written as
\begin{equation}\label{Zbif-rep-F}
   Z_{\text{\sf{bif}}}(\alpha|P',\vec{\mu};P,\vec{\lambda})=b^{-|\vec{\lambda}|-|\vec{\mu}|}
   \prod_{i=1}^{2}\prod_{j=1}^{2}\mathcal{F}(x_{ij},\lambda_{i},\mu_{j}),
\end{equation}
where $x_{ij}=b(Q-P_{i}+P'_{j}-\alpha)$ and 
\begin{equation}\label{Zbif-dual}
    \mathcal{F}(x,\nu,\mu)=\prod_{s\in\nu}(x-l_{\mu}(s)g-(a_{\nu}(s)+1))\prod_{t\in\mu}(x+(l_{\nu}(t)+1)g+a_{\mu}(t)),
\end{equation}
with $g=-b^{2}$. There is another representation  for the function $\mathcal{F}(x,\nu,\mu)$. Let $N$ and $M$ integers such that $N\geq l(\nu)$, $M\geq l(\mu)$ then \cite{Nekrasov:2003rj}
\begin{equation}\label{Zbif-dual-2}
    \mathcal{F}(x,\nu,\mu)=\frac{\prod_{i=1}^{N+1}\prod_{j=1}^{M}\Gamma(x+\mu_{j}-\nu_{i}+(i-j)g)}
    {\prod_{i=1}^{N}\prod_{j=1}^{M+1}\Gamma(x+\mu_{j}-\nu_{i}+(i-j+1)g)}
    \frac{\prod_{i=1}^{N}\Gamma(x+ig)}{\prod_{j=1}^{M}\Gamma(x+(N+1-j)g)}.
\end{equation}
For any $(m,n)\in\lambda$ we find from \eqref{Zbif-dual-2}
\begin{equation}\label{F-FFF}
    \mathcal{F}(x,\lambda,\mu)=\prod_{j=1}^{m}(x-n+jg)_{n}\,
    \frac{\mathcal{F}(x-n,\rho,\mu)\mathcal{F}(x+mg,\nu,\mu)}{\mathcal{F}(x-n+mg,\varnothing,\mu)},
\end{equation}
where $\rho=(\lambda_{1}-n,\dots,\lambda_{m}-n,0,0,\dots)$ and $\nu=(\lambda_{m+1},\lambda_{m+2},\dots)$. Using \eqref{Zbif-rep-F} and \eqref{F-FFF} the identity \eqref{Z-Z-relat} follows.  Similarly, using the property
\begin{equation}
    Z_{\text{\sf{bif}}}(\alpha|P',\vec{\mu};P,\vec{\lambda})=
    Z_{\text{\sf{bif}}}(Q-\alpha|P,\vec{\lambda};P',\vec{\mu})
\end{equation}
one can prove \eqref{Z-Z-relat-2}.
\section{The system of the Integrals of Motion}\label{IM}
One can check that the states $|P\rangle_{\vec{\lambda}}$ are the eigenstates of the following infinite system of mutually commuting  Integrals of Motion
\begin{equation}\label{Integrals}
\begin{aligned}
     &\mathbf{I}_{2}=L_{0}-\frac{c}{24}+2\sum_{k=1}^{\infty}a_{-k}a_{k},\\
     &\mathbf{I}_{3}=\sum_{k=-\infty,k\neq0}^{\infty}a_{-k}L_{k}+2iQ\sum_{k=1}^{\infty}ka_{-k}a_{k}+\frac{1}{3}\sum_{i+j+k=0}a_{i}a_{j}a_{k},\\
     & \begin{multlined}
          \mathbf{I}_{4}=2\sum_{k=1}^{\infty}L_{-k}L_{k}+L_{0}^{2}-\frac{c+2}{12}L_{0}+
          6\sum_{k=-\infty,k\neq0}^{\infty}\sum_{i+j=k}L_{-k}a_{i}a_{j}+12\left(L_{0}-\frac{c}{24}\right)\sum_{k=1}^{\infty}a_{-k}a_{k}+\\
          +6iQ\sum_{k=-\infty,k\neq0}^{\infty}|k|\,a_{-k}L_{k}+2(1-5Q^{2})\sum_{k=1}^{\infty}k^{2}a_{-k}a_{k}+
          6iQ\sum_{i+j+k=0}|k|\,a_{i}a_{j}a_{k}+\hspace*{-5pt}\sum_{i+j+k+l=0}\hspace*{-10pt}:a_{i}a_{j}a_{k}a_{l}:
        \end{multlined}\\
        &\dots\dots\dots\dots\dots\dots\dots\dots\dots\dots\dots\dots\dots\dots\dots\dots\dots\dots\dots\dots\dots     
\end{aligned}
\end{equation}
We note that the system of integrals \eqref{Integrals} can be used to construct the basis $|P\rangle_{\vec{\lambda}}$. Even the integral $\mathbf{I}_{3}$ gives a lot of information. Its eigenstates are our states $|P\rangle_{\vec{\lambda}}$ 
\begin{equation}
    \mathbf{I}_{3}\,|P\rangle_{\vec{\lambda}}=h_{\vec{\lambda}}^{(3)}(P)\,|P\rangle_{\vec{\lambda}},
\end{equation}
with eigenvalues $h_{\vec{\lambda}}^{(3)}(P)$ which are linear functions of the momenta $P$
\begin{equation}\label{Eigenvalue}
   h_{\lambda_{1},\lambda_{2}}^{(3)}(P)=
   h_{\lambda_{1}}^{(3)}(P)+
   h_{\lambda_{2}}^{(3)}(-P),
\end{equation}
where $h_{\lambda}^{(3)}(P)$ is given by 
\begin{equation}\label{eigenvalue}
    h_{\lambda}^{(3)}(P)=i\left(|\lambda|\Bigl(P-\frac{b}{2}\Bigr)+
    \frac{1}{2b}\,\sum_{k}\lambda_{k}(\lambda_{k}+2kb^{2})\right).
\end{equation}
Using \eqref{Eigenvalue} one can show that $\mathbf{I}_{3}$ is not degenerate at levels $1$, $2$ and $3$ while at the level $4$ it has two eigenstates with the same eigenvalue. At the level $5$ it is again non-degenerate. We expect that taking higher integrals $\mathbf{I}_{k}$ the degeneracy will disappear.

We propose that the integrals \eqref{Integrals} are the quantum counterparts of the classical Integrals of Motion $I_{k}=\int G_{k}(x)dx$ 
\begin{equation}\label{classical-integrals}
   \begin{aligned}  
     &G_{2}=u+v^{2},\\
     &G_{3}=uv+v\mathsf{D}v+\frac{1}{3}\,v^{3},\\
     &G_{4}=u^{2}+6uv^{2}+6u\mathsf{D}v+5v_{x}^{2}+6v^{2}\mathsf{D}v+v^{4},\\
     &\hspace*{-2pt}\begin{multlined}
        G_{5}=u^{2}v+\frac{1}{2}\,u\mathsf{D}u+2u_{x}v_{x}+4uv\mathsf{D}v+v^{2}\mathsf{D}u+2uv^{3}+\frac{3}{2}\,v_{x}\mathsf{D}v_{x}
        +\\+3\,vv_{x}^{2}+2v(\mathsf{D}v)^{2}+
        \frac{4}{3}\,v^{3}\mathsf{D}v+\frac{1}{2}\,v^{2}\mathsf{D}v^{2}+\frac{1}{5}\,v^{5},
         \end{multlined}\\
     &\dots\dots\dots\dots\dots\dots\dots\dots\dots\dots\dots\dots\dots\dots\dots\dots\dots\dots\dots\dots\dots\dots.
   \end{aligned}
\end{equation}
where $\mathsf{D}=\frac{d}{dx}\mathsf{H}$ and  $\mathsf{H}$ is the operator of Hilbert transform defined by the principal value integral
\begin{equation}
   \mathsf{H}\,F(x)\overset{\text{def}}{=}\frac{1}{2\pi}\,\dashint_{0}^{2\pi}F(y)\cot\frac{1}{2}(y-x)\,dy,
\end{equation}
The system  \eqref{classical-integrals} is obtained in semiclassical limit $b\rightarrow0$  from the system \eqref{Integrals} via
\begin{equation}
     T\rightarrow -Q^{2}u,\qquad\partial\varphi\rightarrow-iQv,\qquad
     [\;,\,]\rightarrow-\frac{2i\pi}{Q^{2}}\,\{\;,\,\},
\end{equation}
where we have chosen the periodic boundary conditions $T(x+2\pi)=T(x)$, $\partial\varphi(x+2\pi)=\partial\varphi(x)$ with the mode expansion
\begin{equation*}
   T(x)=\sum_{n=-\infty}^{\infty}L_{n}\,e^{-inx}-\frac{c}{24},\qquad
   \partial\varphi(x)=\sum_{n=-\infty,n\neq0}^{\infty}a_{n}\,e^{-inx}.
\end{equation*}
The classical IM $I_{k}=\int G_{k}(x)dx$ are the time conserved quantities  associated with the integrable system  of equations which is known as $\text{Benjamin-Ono}_{\mathbf{\scriptscriptstyle{2}}}$ equation \cite{Lebedev-Radul,Degasperis,Degasperis2}
\begin{equation}\label{Classical-Integrable-Equation}
   \begin{cases}
     u_{t}+vu_{x}+2uv_{x}+\frac{1}{2}v_{xxx}=0,\\
     v_{t}+\frac{u_{x}}{2}+\mathsf{H}v_{xx}+vv_{x}=0,
   \end{cases}
\end{equation}
The system  \eqref{Classical-Integrable-Equation} can be written in a Hamiltonian form 
\begin{equation}
   u_{t}=\{\mathcal{H},u(x)\},\quad v_{t}=\{\mathcal{H},v(x)\},
\end{equation}
with $\mathcal{H}=\int G_{3}(y)dy$ and the ``second'' Hamiltonian structure being of KdV type
\begin{equation*}
   \begin{aligned}
     &\{u(x),u(y)\}=(u(x)+u(y))\,\delta'(x-y)+\frac{1}{2}\,\delta'''(x-y),\\
     &\{v(x),v(y)\}=\frac{1}{2}\delta'(x-y),\quad \{u(x),v(y)\}=0.
   \end{aligned}
\end{equation*}
The classical IM $I_{k}$ form a commutative Poisson bracket algebra with  this Hamiltonian structure.
If we take as a Hamiltonian $\int G_{4}(y)dy$ and further we obtain next representatives of hierarchy related with the integrable system \eqref{Classical-Integrable-Equation}. 

It is also convenient to rewrite the system  \eqref{Classical-Integrable-Equation} in a different form.  Introduce the function
\begin{equation}
\psi=v+iw,
\end{equation}
where $u=w_{x}-w^{2}$ then  the system \eqref{Classical-Integrable-Equation} can be written as an equation for one complex function $\psi$
\begin{equation}\label{BO-equation}
\psi_{t}+\frac{i}{2}\,\psi_{xx}^{\ast}+\psi\psi_{x}+H\,\text{\rm{Re}}\,\psi_{xx}=0,
\end{equation}
which looks like a complexified version of the Benjamin-Ono equation \cite{Benjamin,Ono}. Equation \eqref{BO-equation} simplifies for the function $\psi(x)$ analytic in the upper half plane. In this case it takes a form of the Burgers equation
\begin{equation}\label{Burgers-equation}
\psi_{t}+\frac{i}{2}\,\psi_{xx}+\psi\psi_{x}=0,
\end{equation}
which can be linearized by the Cole-Hopf substitution $\psi=i\,(\log\theta)_{x}$
\begin{equation}\label{Diffusion-equation}
     \theta_{t}+\frac{i}{2}\,\theta_{xx}=0.
\end{equation}
We note that equation \eqref{BO-equation} is equivalent to bidirectional BO equation considered in \cite{Abanov:2009fk}. Namely, let
\begin{equation*}
    \begin{aligned}
      &u_{0}=\frac{1}{2}(\psi+\psi^{\ast}-iH(\psi+\psi^{\ast})),\\
      &u_{1}=\frac{1}{2}(\psi-\psi^{\ast}+iH(\psi+\psi^{\ast})),
    \end{aligned}
\end{equation*}
then equation \eqref{BO-equation} can be rewritten as (compare to (27)--(28) in \cite{Abanov:2009fk})
\begin{equation}\label{2BO-equation}
\psi_{t}+\frac{i}{2}\,\tilde{\psi}_{xx}+\psi\psi_{x}=0,
\end{equation}
where $\psi=u_{0}+u_{1}$ and $\tilde{\psi}=u_{0}-u_{1}$.

\bibliographystyle{MyStyle} 
\bibliography{MyBib}
\end{document}